\documentclass{aa}  
\usepackage{natbib}
\usepackage{silence}
\usepackage[hypertexnames=false,colorlinks=true,linkcolor=blue,citecolor=blue]{hyperref}
\usepackage{comment}
\usepackage{siunitx}
\usepackage{amsmath}
\usepackage{subcaption}
\usepackage{nicefrac}
\usepackage{graphicx}
\usepackage{color}
\definecolor{JDLblue}{rgb}{0.1, 0.3, 0.6}
\usepackage{csvsimple}
\usepackage{ulem}

\usepackage{booktabs,wrapfig}

\usepackage{textcomp,gensymb}
\usepackage{caption} 
\usepackage{tabularx}
\usepackage[table]{xcolor}
\usepackage{collcell}

\newcommand{\diagcell}[1]{%
  \cellcolor{green!20}\ensuremath{#1}
}

\usepackage[bottom]{footmisc}
\usepackage[varg]{txfonts}

\usepackage{listings}
\usepackage{xcolor}

\newcommand{\plx}{\varpi}
\newcommand{\ruwe}{\texttt{RUWE}~}
\newcommand{\truwe}{\texttt{RUWE}}
\newcommand{\gaia}{\textit{Gaia}~}
\newcommand{\tgaia}{\textit{Gaia}}
\newcommand{\di}{\text{d}}
\newcommand{\bigp}{\text{P}}
\newcommand{\prob}[2]{\text{P}(#1 \mid #2)}
\newcommand{\likelihood}[2]{\mathcal{L}(#1 \mid #2)}
\newcommand{\n}[1]{\textbf{[NOTE: #1]}}

\newcommand{\sgt}{\! > \!}
\newcommand{\slt}{\! < \!}

\newcommand{\pbin}{\vec{\theta}}
\newcommand{\data}{\vec{d}}

\begin{document} 

    \title{Mass-dependent multiplicity fraction in low mass stars revealed by \textit{Gaia} astrometry}

   \author{Ben Pennell\inst{1,2}\thanks{Corresponding author: {bepennell@mpia.de}} \and
          Hans-Walter Rix\inst{1} \and
          Kareem El-Badry\inst{3,1} \and
          Sahar Shahaf\inst{1} \and
          Johanna M\"uller-Horn\inst{1,2} \and
          Jiadong Li\inst{1}}

   \institute{
    Max-Planck-Institut für Astronomie, Königstuhl 17, 69117 Heidelberg, Germany
    \and Fakultät für Physik und Astronomie, Universität Heidelberg, Im Neuenheimer Feld 226, 69120 Heidelberg, Germany \and
    Department of Astronomy, California Institute of Technology, Pasadena, CA 91125, USA
}
    \date{Received 22 June 2026 / Accepted **}

\abstract
{Stellar multiplicity at different orbital periods is probed by different techniques: radial velocities at the shortest periods, direct imaging at the longest, and astrometry in between. \textit{Gaia} DR3 provides unprecedented astrometric information to constrain binary populations with periods of tens to thousands of days. Beyond the small fraction of direct orbit solutions, \textit{Gaia} publishes coarser astrometric diagnostics for all objects, such as on-sky accelerations, jerks, or the single-star goodness-of-fit measure, RUWE. We show that these diagnostics together are sensitive to a far wider range of orbital periods than orbit solutions alone. Built on the {\tt Gaiamock} emulator, we develop a forward-modelling framework to constrain rates of multiplicity from \textit{Gaia} data, using all astrometric solution types. We apply this framework to $366~027$ primaries of $0.2$--$0.75\,M_\odot$ between $50$ and $200$\,pc, including vastly more M-dwarfs than previous multiplicity studies. For fiducial models of the period and mass-ratio distributions, the inferred multiplicity fraction is approximately constant at ${\sim}45\%$ from the solar-type regime down to $0.4\,M_\odot$, only then dropping to ${\sim}15\%$ by $0.2\,M_\odot$. 
This mass-multiplicity relationship revises the conventional picture of a monotonically rising mass--multiplicity relation, and aligns the upper M-dwarf regime with solar-type stars rather than placing it on a continuous decline.}
    

   \maketitle
    \nolinenumbers
%

\section{Introduction}

Stellar multiplicity is a fundamental outcome of star formation, and offers one of the few empirical ways of studying it. For example, fragmentation of molecular cloud cores and dynamical processing within young clusters leave their imprint on the occurrence of multiple systems and the distributions their orbital elements \citep[e.g.][]{Tohline2002, Bate2009, Offner+23}.

The underlying physical processes depend on mass, metallicity, and environmental conditions. The multiplicity of field stars reflects the combined outcome of formation and evolution, making the measurement of multiplicity a quantitative probe for both. For instance, the multiplicity fraction as a function of primary mass provides one of the most direct empirical constraints on fragmentation theories \citep{Goodwin+07, Offner+14}.
Dynamical interactions in young clusters process the primordial populations, generally reducing the multiplicity fraction, and, for example, widen loosely bound systems and further harden already tight binaries \citep{Heggie75, Hills75, Kroupa95, Duchene+04, Goodwin+07}.

Binary stars exhibit a wide distribution of orbital separations, spanning several orders of magnitude \citep{DuquennoyMayor91, Raghavan+2010, MoeDiStefano2017}. However, individual observational techniques used to identify binaries are often sensitive to a much narrower range:
Spectroscopic methods, such as radial velocity surveys, are most effective at short periods ($T \lesssim 10^3$\,days), direct imaging and common proper motion searches probe wide separations ($T \gtrsim 10^5$\,days), and astrometric methods cover the intermediate period regime \citep{MoeDiStefano2017}.
As a result, no single technique can capture the full astrophysical picture. Assembling a complete census requires combining all methods with attention to their respective sensitivities.

It is widely accepted that the multiplicity fraction increases with primary mass \citep{Offner+23}.
At the high-mass end, OB-type stars have multiplicity rates approaching unity \citep{SanaEvans11, Sana+2012, MoeDiStefano2017}. 
Solar-type primaries (${\sim}0.8$--$1.2\,M_\odot$) exhibit multiplicity fraction of ${\sim}\,50\%$ \citep[e.g.,][]{MoeDiStefano2017}. 
And, at the lowest stellar masses, empirical multiplicity fractions for M-dwarf primaries ($M_1 {\lesssim}\,0.65\,M_\odot$) yielded values in the range $20$--$30\%$ \citep{DucheneKraus2013, WardDuong+15, Winters+19, Clark+24}. 

However, M-dwarf multiplicity studies have been derived from volume-limited surveys with sample sizes of at most a few thousand objects. These surveys are susceptible to incompleteness from unresolved companions at intermediate separations, a regime that is too wide for spectroscopic detection and too close for imaging. Indeed, 
\cite{Cifuentes+25} recently argued that the multiplicity M-dwarfs has been systematically underestimated. They proposed a corrected multiplicity fraction across M-type primaries closer to $40\%$ based on their candidate unresolved binaries.
Resolving whether the mass--multiplicity relation continues its monotonic decline in the M-type regime, flattens, or exhibits more complex structure requires larger samples with well-understood selection functions. 

Enter the \gaia mission \citep{Gaia} and its third data release \citep[DR3;][]{DR3}, which provides an unprecedented resource for this purpose.
With a baseline of 34 months and several astrometric observations per year for each of its $10^9$ sources, \gaia is sensitive to orbital motion across a broad range of periods, from tens of days to well beyond the observing baseline.

\gaia DR3 publishes astrometric information at several levels of detail.
For all sources, a five-parameter single-star model is fit to the epoch astrometry, and for some sources higher order solutions were fit to model on-sky accelerations and jerks, while a small subset of $168~000$ sources that passed strict quality cuts had their full orbit fit and published \citep{Halbwachs+2023}.
These full orbit solutions have enabled a range of studies, including dormant black hole discoveries \citep{ElBadry+2023BH2, ElBadry+2023BH1, Gaia+2024BH3}, eccentricity statistics \citep{Wu+2025}, and analyses of the companion mass function \citep{Shahaf+19, Shahaf+2024}.

But, the \gaia DR3 suite of astrometric orbits represent only a small fraction of the parameter space; they are confined to ${\sim}10^{2}$--$10^{3}$\,days and favour high signal-to-noise detections. The vast majority of binaries leave subtler astrometric signatures, and were probably unidentified as binary systems by the \gaia pipeline. These systems are nonetheless informative about the underlying population when interpreted statistically.
Extracting population-level constraints from the entire sample requires a forward-modelling framework that predicts the full distribution of \tgaia's astrometric outputs.

In this paper, we develop and apply such a binary population modelling framework using {\tt Gaiamock}, a generative emulator for \tgaia's astrometric pipeline \citep{ElBadry+2024}. 
To illustrate its power in conjunction with \gaia data, we constrain the multiplicity fraction as a function of primary mass for hundreds of thousands of $0.2$--$0.75\,M_\odot$ primaries between 50 and 200\,pc. 
This is a sample orders of magnitude larger than those available to previous companion-counting studies, and spanning the mass range on the main sequence where the mass--multiplicity relation is least well established.

The remainder of this paper is organised as follows: the creation of our sample of nearby low-mass stars will be presented in Section~\ref{sec:gaia}, while also describing the different astrometric signatures of multiplicity available in \tgaia.
Section~\ref{sec:ForwardPopulationModel} outlines the method for how the astrometric signatures can be used to constrain the multiplicity fraction of a sample.
Section~\ref{sec:constraining} will demonstrate the constraining power and self-consistency of this method on synthetic data.
Section~\ref{sec:results} presents our results of applying this machinery to compute the multiplicity fraction as a function of mass for the real sample, the results of which are discussed in Section~\ref{sec:discussion}.
Finally, Section~\ref{sec:summary} will summarise and conclude our results.

\section{\tgaia's astrometric multiplicity signatures} \label{sec:gaia}
In principle, single stars should travel in an effectively straight line on-sky, after correcting for their parallax motion.
\gaia fits a single star model for the sky position $(\alpha, \delta)$, proper motion $(\mu_\alpha,\mu_\delta)$, and parallax $\plx$ of every source \citep{Halbwachs+2023}.
Published alongside those five parameters is a unit weight error that is renormalised to account for systematics arising from colour and magnitude called \truwe, which is constructed to be centred around $1$ \citep{Lindegren+21}.

Spatially unresolved binary stars are astrometrically measured by their centre of light, and for this reason can have inflated \ruwe since their orbital motion can cause their centre of light to deviate from the motion of the centre of mass, making it a good binary diagnostic \citep{Belokurov+20, Penoyre+20, CastroGinard+2024}. 
In \gaia DR3, binary astrometric models were only fit to sources with $\truwe\!>\!1.4$, above which sources are likely to be binaries, since veritable single stars generally cannot get a \ruwe this high \citep{CastroGinard+2024}, aside from errors in measurement such as blending due to nearby sources (Appendix~\ref{app:Spurious}).

Binaries with significant astrometric signals ($\truwe\!>\!1.4$) are fit with higher order solutions with 7, 9, and 12 total parameters.
Binaries with periods up to a couple times the DR3 baseline will have intrinsic on-sky motion that is something like an arc. 
To capture this possibility, these sources are first fit with an additional four parameters corresponding to the on-sky accelerations and their derivatives (jerk, or 9-parameter solution). 
Sources not meeting the quality of fit criteria are then fit with a model only adding the on-sky accelerations (acceleration, or 7-parameter solution). 
If this solution also does not meet the quality criteria, then a model with seven additional parameters which uniquely solves the orbit is fit (full orbit, or 12-parameter solution). These additional parameters are the period $T$, eccentricity $e$, semi-major axis $a$, and four orbital angles $\vec{\xi}$.
For a given primary mass and companion flux contribution, the mass of the companion can be computed.

There are several quality cuts for accepting an orbital solution: the quality-of-fit of the 12-parameter model, and the significance of the fit parallax, semi-major axis, and eccentricity \citep{Halbwachs+2023}.
Sources that are fit with a full orbit solution but don't pass these quality cuts get only their original 5-parameter solution published. The cuts on the full orbit solutions are stringent, resulting in ${\sim}\!168~000$ published full orbit solutions.
Separating the 5-parameter sources into those with low and high \ruwe yields five solution types.
All are assigned the basic 5 parameters $\alpha,\delta,\mu_\alpha,\mu_\delta,\plx$, and the higher order solutions are fit with additional parameters:
\begin{enumerate}
    \item (5-parameter) Low \ruwe ($\truwe < 1.4$)
    \item (5-parameter) High \ruwe ($\truwe > 1.4$)
    \item (7-parameter) Acceleration: $+\dot{\mu_\alpha}, \dot{\mu_\delta}$
    \item (9-parameter) Jerk: $+\dot{\mu_\alpha}, \dot{\mu_\delta},\ddot{\mu_\alpha}, \ddot{\mu_\delta}$
    \item (12-parameter) Full Orbit: $+T,e,a,\vec{\xi}$
\end{enumerate}

\subsection{A sample of low-mass stars within 200pc} \label{sec:query}
\label{sec:astrometry}
\begin{table*}[t!]
    \caption{Summary of the cuts used to define our sample of low mass stars (Section~\ref{sec:query}, Appendix~\ref{app:Sample}). After applying these criteria, the final catalogue contains $366~027$ stars. The number and distribution of sources in each cut is outlined in Table~\ref{tab:SpuriousCuts}}
    \centering
    \small
    \begin{tabularx}{\textwidth}{>{\raggedright\arraybackslash}p{0.24\textwidth}X}
        \hline
        \hline
        Requirement & Applied selection \\
        \hline
        Gaia parent sample & Gaia DR3 sources with published XP spectra; in practice this limits the parent sample to roughly $G\lesssim17.65$ \citep{DR3}. \\
        Distance & $5<\plx<20$\,mas, corresponding to $50$--$200$\,pc. \\
        Faint-companion cut & Best-fit XP binary-model flux ratio $f_2/f_1<10^{-1.5}$ (Appendix~\ref{app:fcut}) \\
        Primary-mass range & $0.2<M_1/M_\odot<0.75$ from the XP-derived parameters \\
        Near main sequence & Within stellar isochrones that represent the main sequence boundary on the CMD (Appendix~\ref{app:Spurious}) \\
        No photometric variability & $\texttt{phot\_variable\_flag} ~!\!= \text{"Variable"}$ (Appendix~\ref{app:Spurious}) \\
        \gaia detects only one source & $\texttt{ipd\_frac\_multi\_peak} <= 1$ \\ 
        Low flux uncertainty & $\texttt{phot\_g\_mean\_flux\_error}/\texttt{phot\_g\_mean\_flux}$ (Appendix~\ref{app:Spurious}) \\
        \hline
    \end{tabularx}
    \label{tab:SelectionCuts}
\end{table*}

\begin{figure}
    \centering
    \includegraphics[width=\linewidth]{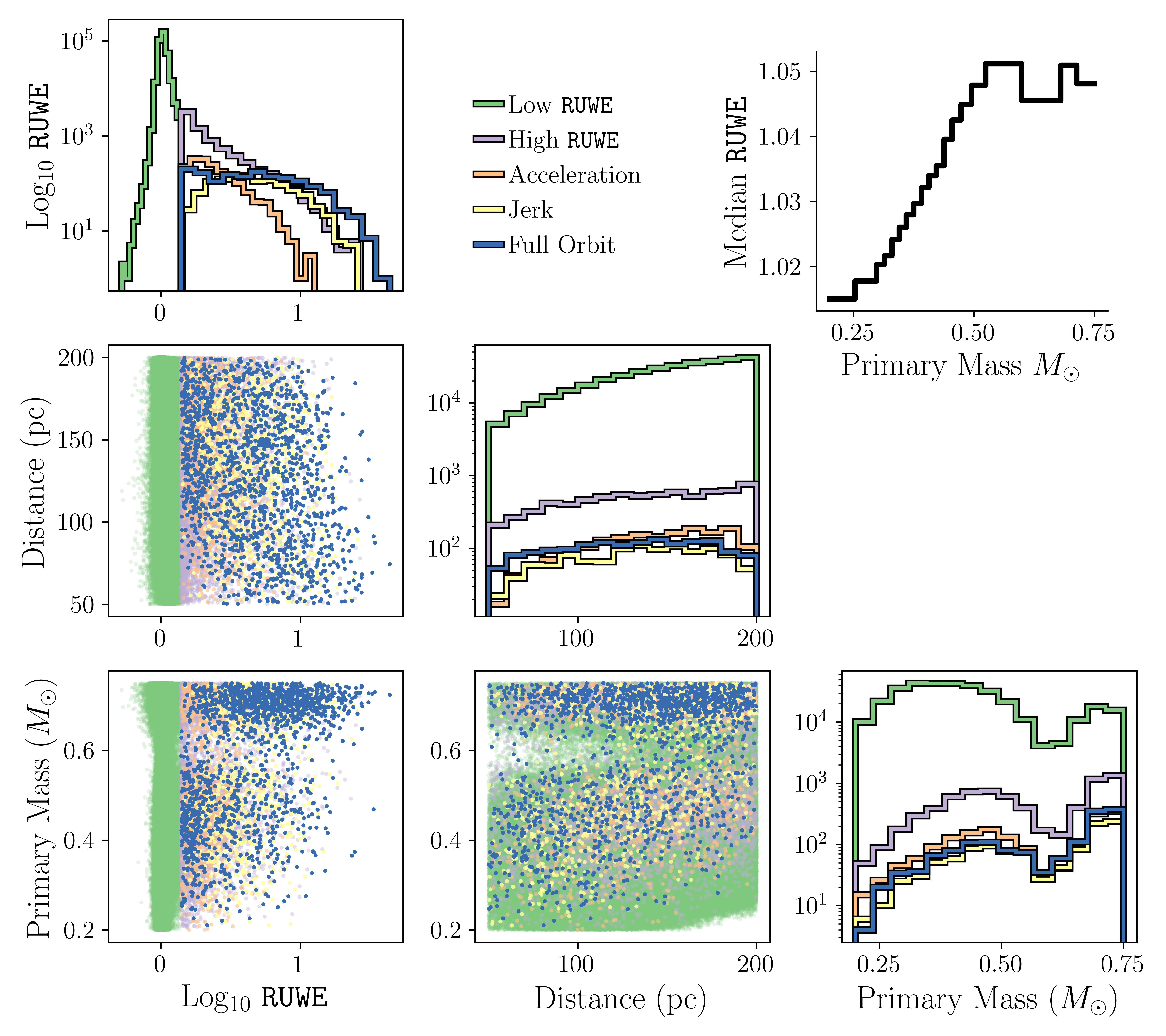}
    \caption{Distributions and covariances of relevant intrinsic parameters in the sample coloured by the astrometric solution type assigned by \gaia for our sample of low mass ($0.2$--$0.75\,M_\odot$) primaries after passing the quality cuts listed in Table~\ref{tab:SelectionCuts}. We aimed to create a sample with only single stars and binaries with companions that contribute negligibly to the flux. This allows us to ignore the companion flux contribution, which makes inference more straightforward. Our sample was created with an emphasis on eliminating spurious sources with inflated \ruwe so as to report a lower limit on the multiplicity fraction. The resulting sample is covered by all astrometric solution types across all primary masses and distances, which shows that we have made reasonable limits to our sample so as to avoid inference on small numbers or bins with zero sources of a particular solution type. In the top right panel we show the median \ruwe as a function of primary mass, which shows an increasing typical astrometric signal strength, which may imply increased multiplicity.}
    \label{fig:SampleCorner}
\end{figure}

\begin{table}[t]
    \caption{Rate and number of each solution type as described in Section~\ref{sec:astrometry} for our sample of nearby low mass primaries.}
    \centering
    \begin{tabular}{l|lllll}
        \hline
        \hline
         Type & $\truwe\!<\!1.4$ & $\truwe\!>\!1.4$ & Accel. & Jerk & Full \\
        \hline
        Rate (\%) & 96.80 & 1.99 & 0.48 & 0.31 & 0.42 \\
        Count & 354320 & 7296 & 1749 & 1121 & 1540 \\
        \hline
    \end{tabular}
    \label{tab:DataSoltypes}
\end{table}

Previous work has focused on taking conservative cuts to the nearest sources to alleviate selection biases.
This is, however, not the only way to construct the sample, only the most straightforward.
Larger samples which extend further and encompass a far larger number of sources can be converted into workable samples if their selection is well-understood.
For this reason, armed with a forward model (Section~\ref{sec:ForwardPopulationModel}, \cite{ElBadry+2024}) we are able to construct samples which reach a far broader set of sources.

To this end, we can construct a sample using all nearby low mass stars found within \gaia DR3, and will construct machinery to model the incidence of companions.
Of course, the process of making such a catalogue requires making many choices.
As will be seen in Section~\ref{sec:results}, we will obtain multiplicity fractions in excess of previous studies, meaning that constructing the sample to minimise the fit multiplicities will strengthen the significance of this result.
With this in mind, when choices must be made when creating the sample, we choose that which results in the lowest multiplicity fraction.
Broadly, the multiplicity fraction will correspond to the rate at which sources get an inflated \truwe, and so we make several cuts to ensure spurious high \ruwe sources are eliminated.
In this section we will explain how the final sample was constructed, and will provide further details in Appendix~\ref{app:Sample}.

The key operating strategy is to create a sample which is well understood and well captured by the forward model.
Small astrometric wobbles can be explained either by a faint low-mass companion, or by a companion of comparable mass and luminosity, which also leads to little photocentre movement.
To get around this degeneracy, we attempt to eliminate all sources with two prominent luminous components through spectroscopic analysis.
This sample cleaning was done with \gaia XP spectra, the DR3 low-resolution (R$\sim$20--100) BP/RP spectrophotometry: time-averaged spectra from \tgaia's Blue and Red Photometers spanning 330--1050\,nm, published for about 220 million sources homogeneously for sources brighter than $G<17.65$ in \tgaia's G-band \citep{DR3,DeAngeli+2023}.
\cite{Li+2025} used these spectra to identify single stars by fitting single-star and binary spectral models to the XP spectra.

First, we fetch all sources between 50 and 200pc which have published XP spectra.
Very few astrometric orbit solutions exist for such stars beyond $200$\,pc, and for this reason we ignore sources beyond this distance.
Sources within $50$\,pc are ignored to have a sample where spatially resolvable companions are far less common (Appendix~\ref{app:cpm}), and leaves the more well-studied sources from the nearby volume-limited studies \citep[e.g][]{WardDuong+15, Winters+19, Clark+24, Gonzalez-Payo2026} as a testset for comparison. 
Additionally, there are issues with astrometry for very nearby ($<10$\,pc) and bright sources, and making this cut alleviates such issues.

We retain sources that were best fit with binary models with a flux ratio less than $10^{-1.5}$, which we've determined retains a fairly pure sample of apparently single stars (and stars with near-invisible companions), as flux ratios below this have little influence on the astrometry (Figure~\ref{fig:fluxcut}).
This flux ratio limit can be translated to a mass ratio limit through isochrone analysis (Appendix~\ref{app:fcut}), which corresponds roughly to a maximum mass ratio $q_\text{max}(M_1)\approx0.35$ (Figure~\ref{fig:MassCutoff}).
Since we've selected sources with no significant companion flux contribution, their colours and absolute magnitudes can more confidently be used to determine the mass and metallicity of the primary \citep{Hwang+2024}.

Armed with mass estimates for every source from \cite{Li+2025}, we select a subsample of sources with primary masses from $0.2$ to $0.75\,M_\odot$.
For the remainder of this paper, the ``primary" corresponds to the present-day brightest component. We assume a small contamination of evolved, more massive companions (Appendix~\ref{app:Spurious}), making the identification of the present-day brightest component as the primary a good assumption.

The $0.2\,M_\odot$ lower limit is chosen as there are few higher order astrometric solutions among primaries of lower mass.
An upper limit of $0.75\,M_\odot$ will bring our studies near the lower extent of the solar-type multiplicity surveys for comparison \citep{Raghavan+2010, Tokovinin14}.

We then make several attempts to clean the sample of spurious detections, each of which is further discussed in Appendix~\ref{app:Spurious}.
First, we use isochrones to trim our sample to the main sequence based on a source's position on the colour-magnitude diagram (CMD).
Then, we eliminate any source flagged as variable by \gaia ($\texttt{phot\_variable\_flag}$).
Then, we eliminate sources where \gaia detects multiple signatures $\texttt{ipd\_frac\_multi\_peak}>1$.
Finally, we remove all sources with high fractional flux uncertainties $\texttt{phot\_g\_mean\_flux\_error}/\texttt{phot\_g\_mean\_flux}$, which appear to be spurious high \ruwe sources due to blending of nearby sources (Figure~\ref{fig:FluxError}).

For these cuts, the removed sources have significantly higher rates of elevated \ruwe than the remaining sample (Table~\ref{tab:SpuriousCuts}), implying these cuts have been both important for multiplicity inference and serve only to deflate the inferred multiplicities.
Spurious sources are also inherently unbiased with primary mass and distance, meaning these cuts will not alter the mass dependence structure (Figure~\ref{fig:CutObjects}).
Generally, the eliminated sources are either expected to be binary-agnostic (variability cut), or biased towards binaries, meaning these cuts result in inferences providing lower bounds on the multiplicity.
Either way, all of these cuts represent sources which cannot be forward modelled by our machinery, and for this reason they must be eliminated from the sample.

It is rightfully a cause of concern that resolved binaries, where each component is identified separately as effectively single, would affect our inferred multiplicities, but restricting our sample to $50\,$pc and our quality cuts make this effect negligible, which is shown in Appendix~\ref{app:cpm}.

Figure~\ref{fig:SampleCorner} shows the distribution of the astrometric solution types as a function of distance and primary mass, and demonstrates that the imposed limits are reasonable as we maintain good coverage of all solution types everywhere.
The number of sources near $0.6\,M_\odot$ is low because the main sequence is quite tight in colour-magnitude space for these primaries (Figure~\ref{fig:isochronecuts}), meaning that it is harder to confirm that no bright companion can be hidden in the spectra, since there are many more combinations of systems which yield similar colour-magnitude positions.
Table~\ref{tab:SelectionCuts} summarises the cuts used to define the sample. The resulting sample has $366~027$ sources. Of these, $318~656$ exist in the conventional M-type ($M_1<0.65\,M_\odot$) regime.

The number of each astrometric solution type in the sample is shown in Table~\ref{tab:DataSoltypes}, and the distributions and covariances between distance, apparent magnitude, primary mass, and \ruwe are shown in Figure~\ref{fig:SampleCorner}.
The top right corner of Figure~\ref{fig:SampleCorner} shows the median \ruwe as a function of primary mass, showing that there are more significant astrometric signatures with higher primary mass.
It is not clear a priori whether this is due to higher multiplicity fraction at higher masses, or higher signal to noise for the brighter sources, which makes astrometric signatures easier to detect, therefore forward modelling is required to convert these rates of detectable multiplicity signatures into a multiplicity fraction. 

\section{A forward population model}
\label{sec:ForwardPopulationModel}

To understand how binary star parameters map to the distribution of astrometric solution types, we need to be able to forward model \gaia observations.
The key algorithmic tool enabling our forward model is {\tt Gaiamock} \citep{ElBadry+2024}, a generative emulator of \gaia DR3's non-single-star (NSS) pipeline. 
It takes in the properties of a binary system and simulates the epoch astrometry that \gaia would record, using the actual \gaia scanning law and per-observation uncertainties for each source. 
From this simulated epoch data it then emulates \tgaia's own solution cascade: computing \ruwe for every source, applying the $\truwe\!>\!1.4$ threshold to trigger higher-order fits, and assigning acceleration, jerk, or full orbit solutions according to \tgaia's quality criteria. 
The result is a predicted probability distribution of observable outcomes for any input binary.
This reproduction of \tgaia's selection function is what makes a quantitative population-level inference possible.

\gaia makes measurements over a grid of CCDs, and \cite{ElBadry+2024} simplified the computation of the astrometric observations by treating each transit as a single observation.
Following the lead of \cite{iorio+26}, we use an updated version of {\tt Gaiamock} which treats each CCD as a separate measurement, and inflates uncertainties based on the sky position of the source. 
\cite{iorio+26} demonstrates how this leads to a better match for the \ruwe distribution of their samples.
For sources that are assigned $\truwe\!>\!1.4$, we use the original version of {\tt Gaiamock} to assign the higher order solution types, as the increased number of observations contribute over an order of magnitude longer orbit solution fitting while not significantly affecting which sources are assigned orbit solutions.

\subsection{Parameter wrangling} \label{sec:Wrangling}
\begin{figure}
    \centering
    \includegraphics[width=\linewidth]{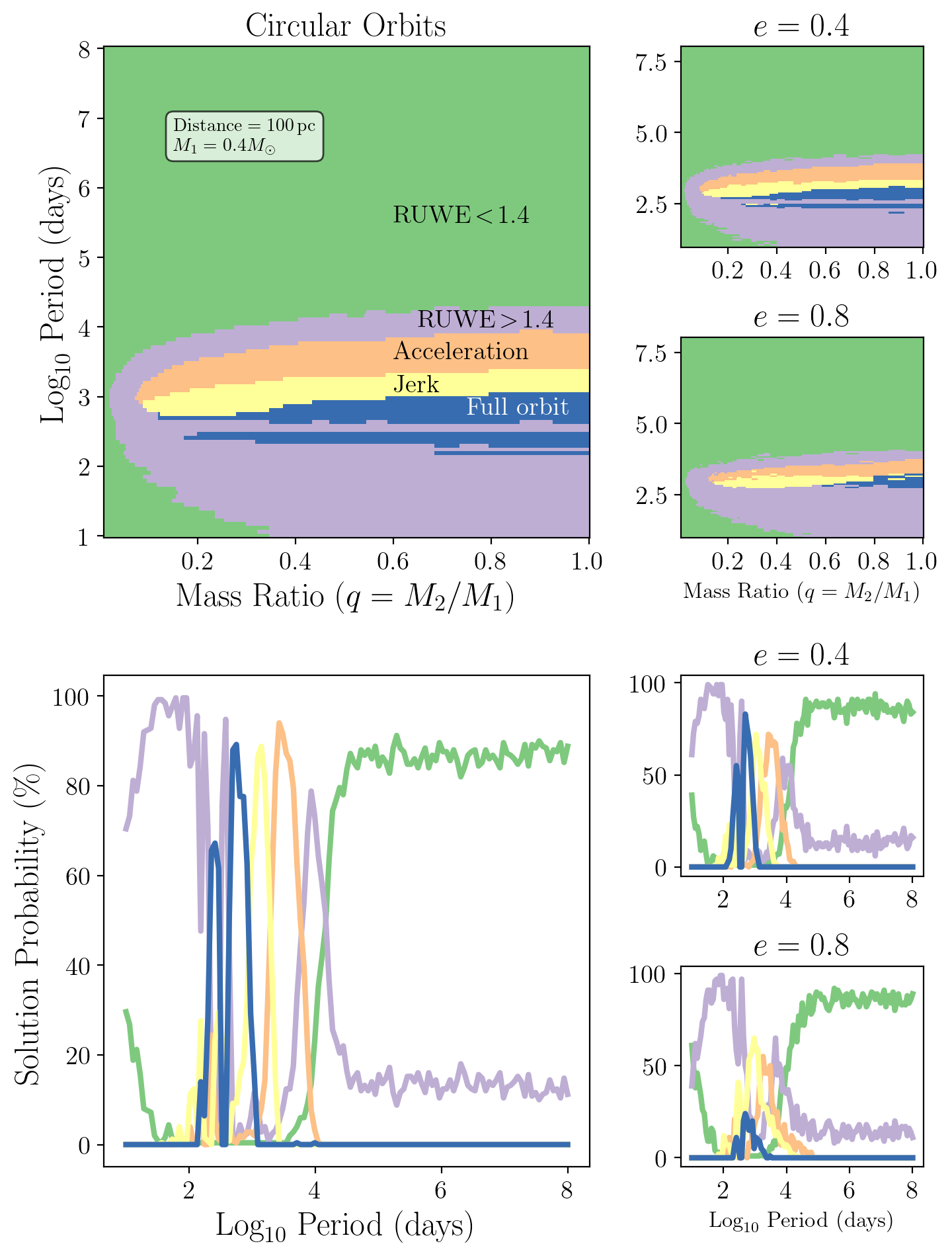}
    \caption{Mapping of basic binary star parameters to Gaia astrometric solution type assigned by {\tt Gaiamock}. Top panel shows the most likely astrometric solution type for a $0.4\,M_\odot$ primary at a distance of $100$\,pc with an invisible companion as a function of mass ratio and period, for different eccentricity regimes. Note the sharp period stratification of solution types, especially for circular orbits. The period range over which Gaia shows astrometric binary stars signatures is vastly wider than the period range for which orbit solutions can be found. Bottom panel represents a vertical slice through the middle of the top panels, showing the rates of each solution type as a function of period when the mass ratio is $0.5$. When the eccentricity is higher, the rates of accelerating and orbit solutions decrease, but occur at higher and lower periods than in the circular case.}
    \label{fig:cube_plot}
\end{figure}

To compute the distribution of some observed \gaia data $\data$, for any particular binary requires fully specifying the orbit
\begin{equation}
    \text{P}(\data) = {\tt Gaiamock}\bigl(T,e,q,M_1,\plx,G,f,\vec{\xi},\alpha,\delta,\mu_\alpha,\mu_\delta\bigr), \label{eq:gaiamock_inputs}
\end{equation}
where T is the orbital period, e is the eccentricity, $M_1$ is the mass of the primary, $q$ is the mass ratio of the secondary to the primary $q=M_2/M_1$, $\plx$ is the parallax, $(\alpha,\delta)$ are the right ascension and declination, $(\mu_\alpha,\mu_\delta)$ are the proper motions, $G$ is the apparent magnitude in the \gaia G-band, and $f$ is the ratio of the secondary to the primary flux in the \gaia G-band. 
Finally, $\vec{\xi}$ represents the four orbital angles required to fully describe the orientation of the orbit, which are conventionally represented as the inclination, longitude of the ascending node, argument of periastron, and time of periastron.

\gaia actually fits and publishes the orbital period, eccentricity, time of periastron and the Thiele-Innes elements \citep{Halbwachs+2023}. The Thiele-Innes elements can then be transformed into the inclination, longitude of the ascending node, argument of periastron and photocentre semi-major axis.

A direct application of {\tt Gaiamock} to population inference is computationally prohibitive: evaluating the likelihood for a single source requires sampling the predicted distribution of solution types over the full space of orbital and nuisance parameters, and a single {\tt Gaiamock} call already takes seconds. For samples of $10^5$--$10^6$ sources, with thousands of likelihood evaluations needed to scan model parameters, on-the-fly simulation is not viable.
The way around this is to precompute a grid of solution-type probabilities spanning all possible binaries, which can be looked up rapidly at inference.
A naive grid over every {\tt Gaiamock} input parameter is itself prohibitive: with twelve parameters (Equation~\ref{eq:gaiamock_inputs}), even a coarse grid of ten points per dimension would require $10^{12}$ {\tt Gaiamock} evaluations.
To reduce this dimensionality, we marginalise over parameters that the astrometric signal does not depend on sensitively, impose distributions on others, and identify a degenerate combination of the remaining physical parameters that further collapses the grid.
The result is a grid of only three variables --- orbital period $T$, eccentricity $e$, and astrometric significance $\Lambda$ defined below --- which can be precomputed once and reused for any sample.
In the rest of this section we derive this dimensionality reduction step by step. 

We can marginalise over a uniform distribution of the orbital angles $\vec{\xi}$ with no loss of generality.
We can also marginalise over sky positions and proper motions by sampling them from the real sample. We assume the flux ratio $f$ is approximately zero.  Appendix~\ref{app:fcut} describes the criteria underlying this selection. This constraint leaves five intrinsic parameters for each system $(T,e,q,M_1,\plx)$. In addition, the apparent magnitude $G$ also plays a role by setting the measurement uncertainties (see further discussion below).

\gaia observes a source's photocentre as it orbits around its centre of mass, which corresponds to a physical semi-major axis $a_\text{phot}$. 
This intrinsic orbit size is projected on the plane of the sky, scaled by the distance which gives a projected orbit angular size $\Theta=a_\text{phot}\plx$. The size of this projected photocentre orbit determines the amplitude of the astrometric signal. Because of this, as an example, \gaia would assign astrometry identically to sources moved further away but with correspondingly increased orbital separations, while maintaining the same angular size, $\Theta$. 

The angular size $\Theta$ is determined by $(T,q,M_1,\plx)$, and can be written as a product of two separate functions, namely, $\Theta = f(T)\lambda(M_1,q,\plx)$. This is because the angular size of an orbit of period $T$ will remain unchanged for all choices of $(M_1,q,\plx)$ as long as they result in the same $\lambda$. We now deduce the functional form of $\lambda$. 

Assuming the companion is invisible, $a_\text{phot}$ is equal to the semimajor axis of the primary's orbit around the centre of mass $a_\text{phot}=a_1$. 
The functional dependence of $\lambda$ follows from Kepler's third law, which relates the orbit's full semimajor axis and the other orbital parameters via
\begin{equation}
    T^2 \propto \frac{(1+q)^2}{q^3} \frac{a_{1}^3}{M_1}. \label{eq:a_tot}
\end{equation}
This is modified from the standard representation of Kepler's third law, since this is written for the motion of the primary around the centre of mass of the system $a_1$, rather than the separation of the two components $a_\text{tot}=\frac{1+q}{q}a_1$.
Now we can cast the projected orbit size in terms of the main quantities of interest:
\begin{equation}
    \Theta =a_1\plx \propto \plx \, T^{2/3}\, M_1^{1/3}\,  q\,  (1+q)^{-2/3} \propto f(T) \, \lambda(M_1,q,\plx),
\end{equation}
which allows us to deduce the form of $\lambda$,
\begin{equation}
    \lambda = \plx M_1^{1/3} \frac{q}{(1+q)^{2/3}}.\label{eq:lambda}
\end{equation}
This is similar to the `Astrometric Mass-Ratio Function'  \citep[denoted $\mathcal{A}$; ][]{Shahaf+19}, which is related to $\lambda$ by $\lambda=\plx M_1^{1/3}\mathcal{A}$. 
This description of $\mathcal{A}$ has proven useful for characterising binary populations \citep{Shahaf+2024, Li2025b, Yamaguchi2025, Rekhi2026}.
For completeness, Appendix~\ref{app:LuminousCompanions} deduces the functional form of $\lambda$ for companions of non-zero flux, following \cite{Shahaf+19}.

\subsection{Astrometric signal significance}

High-\texttt{RUWE} systems in our sample outnumber all Gaia non-single-star solutions combined (Table~\ref{tab:DataSoltypes}). They therefore represent a dominant part of the sample, rather than a small residual category. However, \texttt{RUWE} is not an astrometric orbit parameter. It is a goodness-of-fit statistic for the single-star astrometric model.

It is therefore more natural to discuss these systems in terms of astrometric signal significance. In this view, the relevant quantity is not the absolute angular size of the perturbation, but its size relative to the intrinsic astrometric scatter of the source. This also provides a common language for comparing high-\texttt{RUWE} systems with formal non-single-star solutions: both are manifestations of astrometric signals, but with different levels of information content and model constraint.
Thus astrometric measurements can be described not only by the astrometric signal $\lambda$, but by its corresponding signal-to-noise ratio, $\Lambda=\lambda/\sigma$, which we refer to as the astrometric significance. This is analogous to the significance cuts applied to Gaia DR3 orbital solutions, where accepted orbits were selected using the ratio between the fit photocentre semi-major axis and its uncertainty, $s_\text{orb}=a_0/\sigma_{a_0}$ \citep{Halbwachs+2023}.

For two otherwise identical binaries with the same astrometric perturbation $\lambda$, the astrometric significance changes only through the measurement uncertainty. We treat one system as a reference, $\Lambda_0$. The relation between the target and reference systems signal is 
\begin{equation}
\left(\frac{\Lambda}{\Lambda_0}\right) \simeq \left(\frac{\sigma_{\textsc{g}}}{\sigma_0}\right)^{-1},
\end{equation}
where $\sigma_0$ and ${\sigma_{\textsc{g}}}$ represent the scatter of the reference and target systems, respectively. 
We can therefore write the astrometric significance in terms of the reference scatter level,
\begin{equation}
    \Lambda = \frac{\sigma_0}{\sigma_{\textsc{g}}}\lambda = \frac{\sigma_0}{\sigma(G)}\plx M_1^{1/3} \frac{q}{(1+q)^{2/3}}.\label{eq:AstrometricSignificance}
\end{equation}
We assume that the scatter is a function of the apparent $G$-band magnitude, and neglect colour effects.

The numerical value of $\sigma_0$ is irrelevant for inference: it sets only the overall normalisation of $\Lambda$, and the same $\Lambda$ grid can be reused for any choice. This describes how combinations of $(M_1,q,\plx,G)$ which yield the same $\Lambda$ will provide indistinguishable astrometry, an example of which can be seen in Figure~\ref{fig:LambdaDemonstration}.

\subsection{Solution type distribution} \label{sec:Soltypes}
The parameter reduction described above allows us to compute the distribution of \gaia astrometric signatures using a grid of only three parameters,
\begin{equation}
P(\vec{d}) = {\tt Gaiamock}\bigl(T,e,\Lambda\bigr),
\end{equation}
where the source-dependent quantities $(M_1,\plx,G,q)$ enter only through their contribution to $\Lambda$. At inference, the expected astrometric response of each source can therefore be obtained from the precomputed grid at the corresponding value of $(T,e,\Lambda)$.

In practice, we use this grid to compute $\prob{k}{T,e,\Lambda}$, the probability that \gaia assigns solution type $k$ to a binary with period $T$, eccentricity $e$, and astrometric significance $\Lambda$. To express this selection function in terms of binary parameters, we convert $\Lambda$ to companion mass ratio. 
This conversion is set by the source properties $(M_1,\plx,G)$: the primary mass and parallax determine the angular size of the photocentre orbit, while the apparent magnitude sets the astrometric noise. Thus, for each source, the same grid can be read as a selection function in $(T,e,q)$-space.

Figure~\ref{fig:cube_plot} illustrates this selection function for a representative M-type primary. The top panels show the most likely solution type across the period--mass-ratio plane, for several eccentricities. The bottom panels show the corresponding solution-type probabilities at fixed mass ratio, $q=0.5$. The main ordering is by orbital period, because the period determines the form of the photocentre motion over the \gaia observing baseline. Mass ratio and eccentricity then modify this structure: mass ratio changes the signal amplitude, while eccentricity changes how the orbital motion is sampled in time.

The previous step gives the selection function $\prob{k}{T,e,q}$ for binaries with specified orbital parameters. In the population model, these parameters are not known for each source. They are treated as latent variables drawn from an underlying distribution,
\begin{equation}
\prob{T,e,q}{\pbin},
\end{equation}
where $\pbin$ denotes the parameters of the binary-population model. A choice of $\pbin$ therefore specifies how often binaries occupy different regions of $(T,e,q)$-space.
Combining this population model with the \gaia selection function predicts the distribution of astrometric solution types in the sample. The population parameters can then be constrained by comparing this prediction to the observed data. In this paper, the data $\data$ are the solution-type labels  for all sources in the sample (denoted $\vec{k}$), therefore, the corresponding likelihood is $\prob{\vec{k}}{\pbin}$.

For a dataset of $N$ independent objects, we can compute the probability of reproducing the entire ensemble of observations object-wise:
\begin{equation}
    \prob{\data}{\pbin} = \prod_i^N \prob{\data_i}{\pbin}. \label{eq:totallikelihood}
\end{equation}
In most applications, the probability $\prob{\data_i}{\pbin}$ also uses an object's intrinsic properties $\vec{\eta}$, such as in our case we use each source's assigned primary mass and parallax at inference.
For this reason Equation~\ref{eq:totallikelihood} should in principle be written with this dependence $\prob{\data_i}{\pbin,\vec{\eta}}$, but for clarity we omit this implicit dependence in the notation for the rest of the paper.

We use {\tt Gaiamock} to compute the probability $\prob{\data_i}{\pbin}$ of reproducing a single object's data $\data_i$. 
Through forward modelling of the selection function, this implicitly corrects for the incompleteness of a volume-limited sample, as further or fainter sources will simply contribute less to the rates of higher order solution types. 
Further, since we focus on learning the population properties, rather than those of the individual sources, we do not include the uncertainty in the assigned parallax or mass of the sources. 
In our sample, our cuts to the main sequence (Appendix~\ref{app:Spurious}) and limit to 200\,pc result in low uncertainties for both properties. 
The \texttt{parallax\_over\_error} reported in \tgaia's source table for each source is already greater than 10, and the dynamical masses have fractional uncertainties of around $10\%$ for this region on the main sequence \citep{Hwang+2024}.
Provided these uncertainties are unbiased and small relative to the bin widths over which we infer MF, they will broaden but not shift the recovered MF(M$_1$) relation, and our percent-level inference is robust to them.
Finally, as can be seen in Figure~\ref{fig:CountDependence}, percent level biases already exist in the inference.

\subsection{An objective function}
An objective scalar function that can be extremised is required to constrain the values of model parameters, whether through analytical, numerical or machine learning methods. 
We start with the probability of reproducing data $\data_i$ for individual objects $i$ given the binary model $\prob{\data_i}{\pbin}$. 
However, we cannot compute this probability directly in {\tt Gaiamock}.
By choosing models for the distributions of the orbital parameters, we can compute $\prob{T,e,q}{\pbin}$, which converts model parameters $\pbin$ into physically relevant variables period $T$, eccentricity $e$, and mass ratio $q$. 
Then, we can use {\tt Gaiamock} to forward model our data $\prob{\data_i}{T,e,q}$ using these binary parameters. 
If we do this for all combinations of $(T,e,q)$, we obtain the probability $\prob{\data_i}{\pbin}$. In the limit of infinite resolution, can represent $\prob{\data_i}{\pbin}$ as an integral over the range of $(T,e,q)$,
\begin{equation}
    \prob{\data_i}{\pbin} = \int\!\di T~\di e~\di q~\prob{\data_i}{T,e,q}~ \prob{T,e,q}{\pbin}. \label{eq:prob}
\end{equation}
Our objective function is the combined probability $\prob{\data_i}{\pbin}$ for each source $i$, which is expressed naturally by logarithmic summation:
\begin{equation}
    \likelihood{\data}{\pbin} = \sum_i^N \log(\prob{\data_i}{\pbin}). \label{eq:log_likelihood}
\end{equation}

\subsection{Placing constraints with solution types}
\begin{figure*}[t]
    \centering
    \includegraphics[width=\textwidth]{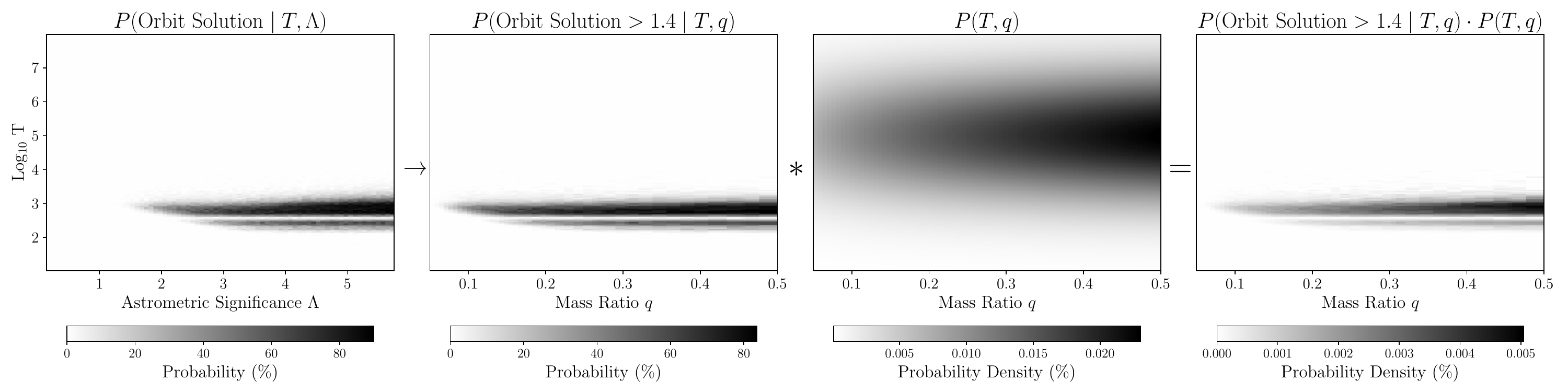}
    \caption{Schematic illustration of our forward modelling approach. We show how the likelihood of getting a full orbit solution is calculated for a $0.38\,M_\odot$ source at $\plx=8$\,mas. From left to right: first, we pre-compute the probability of getting a full orbit solution for a all combinations of the orbital period $T$ and the Astrometric Significance $\Lambda=\Lambda(M_1,q,\plx,G)$ (Equation~\ref{eq:AstrometricSignificance}). Then, for this particular source, we use its primary mass, distance, and apparent magnitude with Equation~\ref{eq:AstrometricSignificance} to convert from $\Lambda$ to mass ratio $q$, yielding $\prob{\text{Orbit Solution}}{T,q}$. Then, we consider the model for the $T$ and $q$ probability distributions under consideration. For this example we used a log-normal period and power law mass ratio distribution $\bigp(q)\propto q^{0.5}$. Finally, we compute the joined probability through element-wise multiplication $\prob{\text{Orbit Solution}}{T,q}\cdot\bigp(T,q)$. The likelihood is computed by summing up the resulting grid. For this example, where the solution type was the 12-parameter full orbit solution, this results in a roughly $2.5\%$ chance for this source to be correctly assigned the right solution type given the chosen model. One can see that if the period distribution were more centred around the \gaia baseline, or more biased towards higher mass companions, this probability would increase.}
    \label{fig:demonstration}
\end{figure*}

From Equations~\ref{eq:prob} and \ref{eq:log_likelihood}, we can write the likelihood of reproducing the ensemble of solution types $\vec{k}$ in the dataset as
\begin{equation}
    \likelihood{\vec{k}}{\pbin} = \sum_i \log\biggl(\int\!\di T~\di e~\di q~\prob{k_i}{T,e,q}~\prob{T,e,q}{\pbin}\biggr).
\end{equation}
A random sample of sources in \gaia is composed of single and binary stars.
Single stars are in principle not of interest for studying the distribution of binary properties, however most binaries are astrometrically indistinguishable from single stars, and the contribution of binaries to the astrometric signals in this regime is still information and therefore should not be discarded.
Ignoring the astrometrically uninteresting, while expediting possible inference, rejects the majority of the available information.
Modelling the occurrence of the single stars is also needed to determine the multiplicity rate.

Notably, it is not logical to compute a probability for a single star using Equation~\ref{eq:prob} because assigning binary parameter distributions to a single star is meaningless. 
We therefore split the probability of reproducing the solution types $\vec{k}$ into single-star (ss) and non-single-star (nss) contributions and treat them as a mixture model, adding an explicit single-star term:
\begin{equation}
    \prob{k_i}{\pbin} = \prob{k_i}{\pbin,\text{ss}}\prob{\text{ss}}{\pbin} + \prob{k_i}{\pbin,\text{nss}}\prob{\text{nss}}{\pbin}.
\end{equation}
There are four possible solution types in \tgaia, which we'll label by the number of parameters: $k\in(5,7,9,12)$. However, the 5-parameter sources come in two flavours: those with $\truwe\slt1.4$ which are indistinguishable from single stars, and those with $\truwe\sgt1.4$ which have this significant astrometric signal but didn't pass the higher order solutions quality cuts. 
We will label objects with $\truwe\slt1.4$ as $k_i=0$, and leave $k_i=5$ to only objects with $\truwe>1.4$. We can also rewrite $\prob{\text{nss}}{\pbin}$ as the multiplicity fraction $\text{MF}$, and $\prob{\text{ss}}{\pbin}$ as its complement, so that
\begin{equation}
    \prob{k_i}{\pbin} = (1-\text{MF})\delta(k_i) + \text{MF}\cdot\prob{k_i}{\pbin,\text{nss}},
\end{equation}
where $\delta(k_i)$ is a Kronecker delta on the solution-type label, and $\prob{k_i}{\pbin,\text{nss}}$ is described by Equation~\ref{eq:prob}.

For the remainder of this paper we will only constrain the multiplicity fraction, leaving the simplest case where $\pbin$ is fixed at fiducial values.
We also assume that the population contains only single stars and binaries with stellar or sub-stellar companions; that spurious detections (Appendix~\ref{app:Spurious}), higher-order systems, or other kinds of companions contribute negligibly. An advantage of focusing on smaller primaries is that this is a better approximation than for massive primaries, which have higher rates of being triples or higher-order systems.

For an individual object, the entire process goes as follows: Use {\tt Gaiamock} to compute the solution type probability distribution $\prob{\text{k}}{T,\Lambda}$ that spans all possible values of $T$ and $\Lambda$, marginalising over some fiducial eccentricity distribution.
Then, use the object's intrinsic $(M,\plx,G)$ to convert the grid from $\Lambda$ space into $q$ space (Equation~\ref{eq:AstrometricSignificance}).
Then, a choice must be made on a particular model of the orbital parameter distributions $\prob{T,q}{\pbin}$ computed along a grid of $T$ and $q$. 
If the grids $\prob{\text{k}}{T,q}$ and $\prob{T,q}{\pbin}$ are the same size, then the total probability expressed as an integral in Equation~\ref{eq:prob} can be computed by element-wise multiplication and summation of the resulting grid which yields the likelihood.
In this work, we compute the likelihood over a fine grid of the entire range of multiplicity fractions.
This strategy, built on creating grids of probability density as a coarse approximation of integration can be vectorised and parallelised to make inference rapid, as vectorised element-wise multiplication can be computed faster than a single orbit solution fit in {\tt Gaiamock}.
Figure~\ref{fig:demonstration} gives a visual representation of the implementation of this computation scheme.

\section{Validation on synthetic data} \label{sec:constraining}
We now demonstrate that this method is self-consistent, by showing that a mass-multiplicity relationship can be accurately and precisely recovered in synthetic datasets.

\subsection{Creating a synthetic sample} \label{sec:synthdata}
Synthetic objects are created by first sampling sky positions, proper motions, primary masses and apparent magnitudes randomly from a real catalogue.
This preserves the spatial, mass, and magnitude distributions of objects from the real sample, alleviating the need to impose models for them.
For the remainder of this paper, we will use the sample of objects described in Section~\ref{sec:query}.
Then, some fraction of objects are assigned companions with random orbital angles. 
The orbital parameters: periods, mass ratios, and eccentricities are assigned with fiducial models.
These objects are then forward-modelled using the generative forward-model described in Section~\ref{sec:ForwardPopulationModel} to determine their astrometry, summary statistics, and assigned solutions. 
For this paper, we choose empirical fiducial models for the orbital parameters, and report the multiplicity fraction conditioned on these parameters.

It is generally expected that orbital periods follow something like a log-normal distribution \citep{Opik1924, DuquennoyMayor91, MoeDiStefano2017, Raghavan+2010, DucheneKraus2013}, which can be parametrised by the location of their peak $T_\mu$, and their spread $T_\sigma$:
\begin{equation}
    \prob{T}{T_\mu,T_\sigma} ~=~ \frac{1}{\sqrt{2\pi T_\sigma^2}}~~\text{exp}\biggl(-\frac{\bigl(\log{T}-T_\mu\bigr)^2}{2T_\sigma^2}\biggr). \label{eq:log-normal}
\end{equation}
\cite{DucheneKraus2013}  and \cite{Clark+24} describe an empirically fit log-normal peaking in log-days at $T_\mu\sim4$ with a width of $T_\sigma\sim1.3$ for their sample of low mass stars, which we choose for our synthetic data. 

Mass ratio distributions are often parametrised with a power law $\bigp(q)\propto q^\gamma$. Low mass stars are observed to follow a mass ratio distribution that is either flat or a power law with a modest index \citep[$\gamma\sim0.5$][]{Delfosse2004, DucheneKraus2013, ReggianiMeyer2013}. 
Further, the mass ratio distribution may be dependent on the mass of the primary and the separation of the secondary, where smaller primaries and closer separations can increase the index \cite{DucheneKraus2013}.
Simulations have found that lower mass primaries prefer a flatter mass ratio distribution than their solar mass counterparts \citep{Bate2009, Bate2012}.
For simplicity, we take a flat mass ratio distribution for all regimes, which will result in the lowest inferred multiplicity fraction.

As can be seen in Figure~\ref{fig:cube_plot}, the eccentricity also has a meaningful impact on the distribution of solution types, which is due to the orbit coverage of the observations at higher periods. 
Larger astrometric signatures appear near periapsis due to the faster orbital speed, which allows high-period and high-eccentricity objects to obtain high order solutions otherwise not possible with a circular orbit. 
However, the majority of time is spent travelling slower near apoapsis, resulting in an overall lower rate of high order solution assignment by \tgaia. 
Some care needs to be taken in the selection of eccentricity distributions, as use of an incorrect eccentricity model can therefore induce biases of several percent in the fit multiplicity fractions (Appendix~\ref{app:Robustness}).

From energy equipartition considerations, binary stars ought to settle into a `thermal' distribution $\text{P}(e)\!=\!2e$ \citep{Jeans1919, Ambartsumian1937}.
This is in a way the `canonical' distribution of eccentricities often used in the absence of further information.

However, the eccentricity may peak at lower eccentricities for lower periods, and the near completeness of the \gaia orbit solutions for nearby solar stars shows that they follow something closer to a normal distribution peaked at $e\!=\!0.4$ \citep{Wu+2025} for binaries near the DR3 baseline, which is also found in surveys of visual binaries \citep{DuquennoyMayor91, Tokovinin2016}.
This distribution takes the form of a Rayleigh distribution, \cite{Wu+2025} provided several potential physical explanations, but these need not continue to dominate for higher periods where dynamical interactions become dominant and the assumptions for the thermal distribution become closer to valid, as is observed in wide binaries \citep[e.g.][]{Tokovinin2016, Tokovinin2020, Hwang+2022}.
For this reason, we propose a `turnover' model, where the eccentricity model smoothly transitions between a normal and thermal distributions as the period increases. The model is parametrized as
\begin{equation}
    \bigp_{\text{turnover}}(e,T) = \bigl(1-w(T)\bigr)\bigp_{\text{Rayleigh}}(e) + w(T)\bigp_{\text{Thermal}}(e),\label{eq:eccentricity}
\end{equation}
where $w(T)\in(0,1)$ is a function of period that sets where and how quickly the model moves from Rayleigh to thermal.
We choose $w(T)$ to be the cumulative distribution function of a normal distribution centred at $\log{T} =3.5$ with a width of 0.5, such that it roughly transitions between the Rayleigh and Thermal distributions between $10^3$ and $10^4$\,days.

A summary of the choices for the distributions of the orbital parameters is found in Table~\ref{tab:choices}.
We show in Figure~\ref{fig:NSSComparison} that each of these models provide reasonable fits to the observed distributions in the objects assigned orbit solutions in the observed sample, which suggests that the model uncertainty will not be large.
In Appendix~\ref{app:Robustness}, we show the fit multiplicity fractions for different model choices, where different models can provide very different multiplicity fractions.
However, the agreement with the observed orbit solutions suggests that we can expect the choices of period, mass ratio and eccentricity to correspond to not more than $\mathcal{O}(\%)$ biases, which is already the scale of the biases from inference (Figure~\ref{fig:CountDependence}).
\begin{table}[t]
    \caption{Orbital parameter model choices chosen to reasonably reflect reality. Their match to the sample can be seen in Figure~\ref{fig:NSSComparison}.}
    \begin{tabular}{l|l}
    \hline
    \hline
     $\bigp(T)$ & log-normal (Equation~\ref{eq:log-normal}) with $T_\mu=4$, $T_\sigma=1.3$ \\
     $\bigp(e)$ & `turnover' (Equation~\ref{eq:eccentricity}) \\
     $\bigp(q)$ & Uniform \\
    \hline
    \end{tabular}
    \label{tab:choices}
\end{table}

\subsection{Results on synthetic data} \label{sec:synthresults}
\begin{figure}
    \centering
    \includegraphics[width=\linewidth]{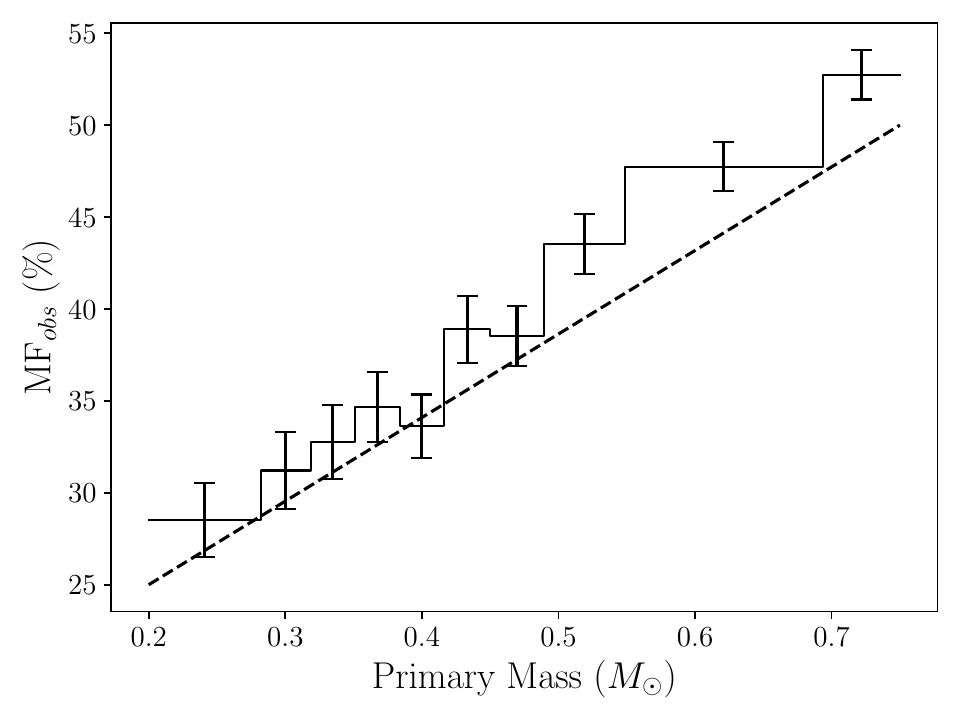}
    \caption{Verification of our modelling approach on synthetic data, where the input multiplicity fractions (dashed line) are well recovered at inference (steps). Synthetic data for $250~000$ objects were constructed through forward modelling with {\tt Gaiamock} to represent the distances, directions and magnitudes of the actual sample from Section~\ref{sec:query}. Fiducial models for the period, eccentricity and mass ratio distributions were chosen as described in Section~\ref{sec:constraining}. The synthetic sample was split into ten bins with equal number of sources by primary mass. Our machinery then assigns the requisite multiplicity fraction for each sub sample to reproduce the rates of each astrometric solution types. This shows the effectiveness of the inference scheme. The percent-level bias noted by bootstrapping in Figure~\ref{fig:CountDependence} can also be seen here, showing the limits of our inference.} 
    \label{fig:binarity_example}
\end{figure}

For the remainder of this paper, we estimate inference uncertainties from the Cramér-Rao lower bound.
In other words, the lower limit on the uncertainty is estimated from the second derivative at maximum likelihood with respect to the parameter.
This is a good approximation of the real uncertainty found by bootstrapping (Figure~\ref{fig:CountDependence}); the agreement between Cramér-Rao and bootstrap scatter validates the error bars we report.
Evidently this does not capture the uncertainty due to model selection, since we have no reasonable prior to place on input models we can't mindfully assign a particular uncertainty for this.

We compute a grid for the probability of getting each solution type $\prob{k}{T,\Lambda}$ for a grid of 150 evenly spaced grid points in log period space $1<\log{T}<8$ and 75 grid points sampled evenly in $\Lambda^{1/4}$ space such that the $\Lambda$ spacing covers the mass ratios $0.05<q<0.5$ to capture every possible companion. 
The power law distribution sampling of $\Lambda$ is done in an attempt to better assign resources since most objects take on lower values of $\Lambda$. 
It was found that changing the size of the grid made little effect on the inference if it is larger than 10 bins in each dimension. 
At each point in the grid, 250 binaries are randomly generated from the marginalised parameters and have their astrometry and solution type assigned using {\tt Gaiamock} to get the rate of each solution type in that bin. 
For this work, the eccentricity is also marginalised over, using the `turnover' model (Equation~\ref{eq:eccentricity}). 

To demonstrate the efficacy of revealing a mass-multiplicity relationship, we construct a synthetic dataset of $\sim 250~000$ objects with the same method as above and we introduce a variable multiplicity fraction that increases linearly with mass from $25\%$ at $0.2\,M_\odot$ to $50\%$ at $0.75\,M_\odot$, and in Figure~\ref{fig:binarity_example} we show that this relationship is well captured, with only $\mathcal{O}(\%)$ errors and biases throughout.
We compute the multiplicity fraction (MF)
\begin{equation}
    \text{MF} = \frac{\text{\#Binary}}{\text{\#Binary} + \text{\#Single}} \label{eq:mf}
\end{equation}
restricted to mass ratios that we retain in our sample (Appendix~\ref{app:fcut}), which we label the observed multiplicity fraction $\text{MF}_{obs}=\text{MF}(q<q_\text{max})$.

This emulation scheme has proven to be self consistent, and the real advantage is in computing performance.
The inference is done on the timescale of minutes, and since computing the probability distributions for each object requires hundreds of {\tt Gaiamock} computations, this means that the dimensionality reducing scheme to create a cached grid has yielded inference speed up by a factor of hundreds or thousands compared to on-the-fly calculations.
Furthermore, this becomes more profitable as the sample size increases since the cached grid is the same regardless of sample size.

\subsection{Placing joint constraints} \label{sec:modelling}
We have demonstrated on mock data that we can place meaningful constraints on the multiplicity fraction, under the assumption that we understand the uncertainties and systematics in \gaia well enough for {\tt Gaiamock} to reproduce the \ruwe and astrometric cascade pipeline, and that outliers and unaccounted for systems contribute minimally.

\cite{ElBadry+2024} showed that {\tt Gaiamock} can broadly compute the correct rate of orbital solutions for simulated observations of the entire Milky Way. 
In our synthetic datasets, the rates of orbit solutions were significantly higher than those of the real sample, which indicates the existence of uncertainties or biases unaccounted for in our forward modelling framework. 

Our sample features sources that are faint, the majority of which are fainter than $G\!=\!14$, beyond which uncertainties increase significantly \citep{Lindegren+21, Holl+2023}. Furthermore, \gaia computes \ruwe by an empirical scaling of the unit weight error based on the sky position and colour of the source, which centres the \ruwe distribution around $1$, but can introduce errors for very red objects \citep{Lindegren+21}. 
So, while the uncertainties contained within \gaia may be well enough understood on average across many regimes, it is possible that when the objects get sufficiently red, or dim, we start to lose modelling precision. 
It was also not validated in \cite{ElBadry+2024} whether {\tt Gaiamock} reproduces the rate of acceleration and jerk solutions accordingly.

For this reason, as a demonstration for the remainder of this paper, we compute only the multiplicity fraction, as \cite{iorio+26} demonstrated that the \ruwe distribution can be well captured, providing us one degree of freedom.
For this paper, we make no attempt at jointly inferring the distributions of orbital parameters.
With this all in mind, we anticipate that future data releases from \tgaia, in particular the per-epoch astrometric data available in its next data release, will provide the opportunity to model the distribution of astrometric signals as we will have a more complete understanding of the raw data.
Furthermore, we will have the ability to fit our own astrometric solutions and outlier treatments, which will minimise the difference between forward models and measurable reality.

\section{Application: multiplicity in low-mass stars} \label{sec:results}
We now proceed to a first application of our approach: we estimate the multiplicity fraction for low-mass primaries on the main sequence, as a function of their mass. 
For this, we can use our sample of $366~027$ nearby $0.2$--$0.75\,M_\odot$ primaries outlined in Section~\ref{sec:query}, which allows us to look at $\text{MF}(M_1)$ (Equation~\ref{eq:mf}) in unprecedented detail.
Following what was outlined in Figure~\ref{fig:binarity_example}, we place our sample into 15 bins of equal number of sources by $M_1$, and constrain the MF in each bin for a set of fiducial models for the distribution of periods, eccentricities and mass ratios (Table~\ref{tab:choices}).
For inference, we use the same constructed grid as outlined in Section~\ref{sec:synthresults}.

\begin{figure}
    \centering
    \includegraphics[width=\linewidth]{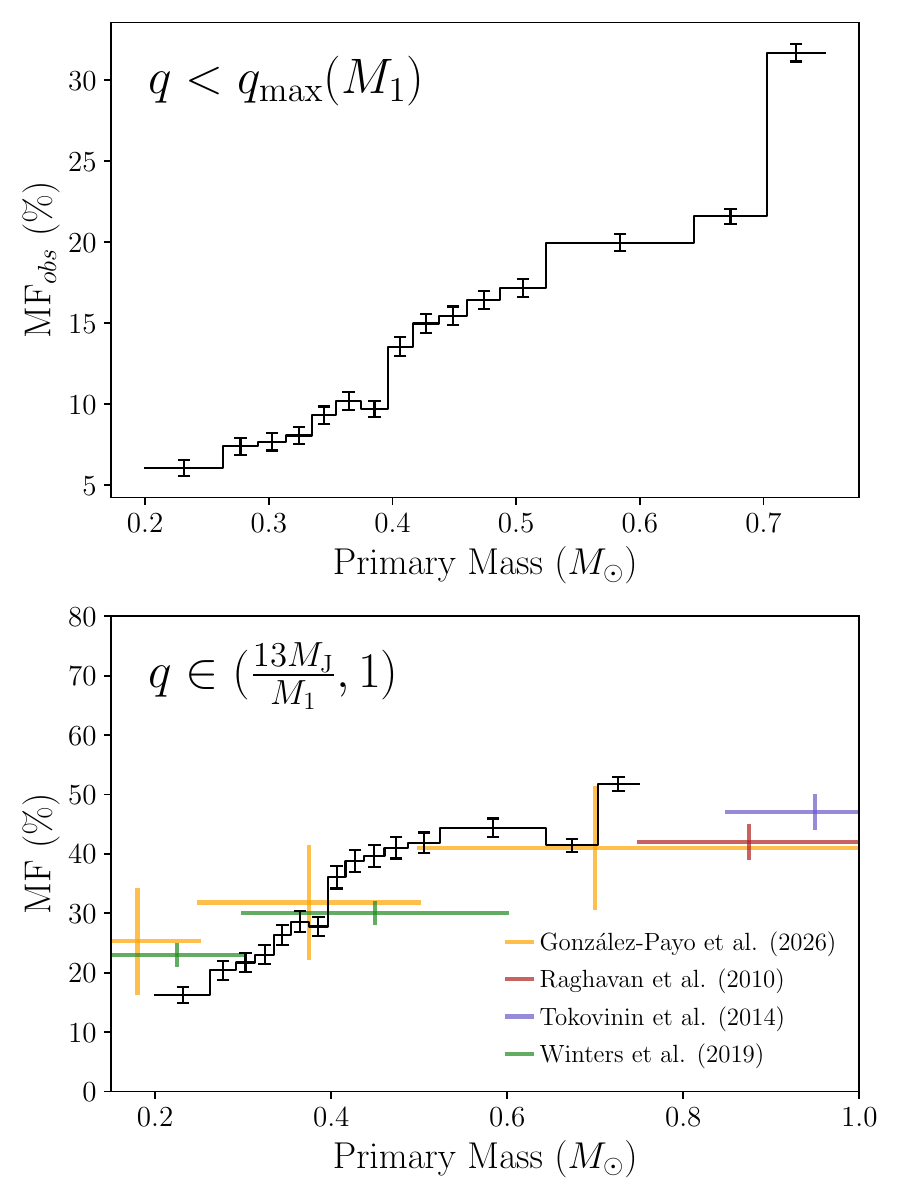}
    \caption{Lower limit multiplicity fractions as a function of primary mass for low-mass stars. These estimates used the fiducial models for the period, eccentricity, and mass ratio distributions: the period follows a log-normal distribution peaking at $10^4$ days with a width of 1.3, the eccentricity follows the empirical "turnover" model (Equation~\ref{eq:eccentricity}), and the mass ratios follow a uniform distribution. These fiducial models well reproduce the distributions of these parameters in the objects assigned orbit solutions (Figure~\ref{fig:NSSComparison}). Our sample was restricted to objects where spectral analysis restricts the flux contribution of a possible companion (Section~\ref{sec:query}), which corresponds to a maximum mass ratio $q_{max}(M_1,[M/H])$ (Appendix~\ref{app:fcut}), and the multiplicity fraction we infer MF$_\text{obs}$ is constrained to objects below this mass ratio (top panel). The bottom panel shows the extrapolation to the multiplicity fraction, MF, across all mass ratios, assuming the mass ratio distribution remains uniform. We further plot literature values \citep{Raghavan+2010, Tokovinin14, Winters+19} as adapted from Figure 1 in \cite{Offner+23} and \cite{Gonzalez-Payo2026} as crosses.}
    \label{fig:results}
\end{figure}

In the top panel of Figure~\ref{fig:results}, we show the inferred multiplicity fraction in each bin $\text{MF}_{obs}=\text{MF}(q<q_\text{max})$.
This multiplicity fraction corresponds to the fraction of objects with a companion with a mass ratio less than the limits set by the flux limit in Appendix~\ref{app:fcut}, which is generally $q_\text{max}\sim0.35$, but is a function of primary mass and metallicity.
The mass-multiplicity relationship shows a fairly flat distribution above $0.4\,M_\odot$, with a decline by a factor of three to $0.2\,M_\odot$.

We cannot directly measure the occurrence of companions $q>q_\text{max}$, so we extrapolate to a multiplicity fraction across all mass ratios by assuming the distribution remains flat up to $q=1$.
As pointed out in \cite{MoeDiStefano2017}, solar mass stars appear to have an excess of twins ($q>0.95$), but this isn't as well known for lower mass primaries, and so we ignore their contribution.
Both the assumption of a flat distribution and lack of twin excess provide the minimum total multiplicity fraction.
Finally, we remove from this extrapolation any region of $q$-space which would be occupied by conventionally planetary companions $M_2\lesssim13M_\text{Jup}$ making the final mass ratio space $q\in(\frac{13M_\text{Jup}}{M_1},1)$, such that we report a multiplicity of stellar (and brown dwarf) companions, to make our reported multiplicities more directly comparable to other studies.
This extrapolation places the multiplicity as increasing from ${\sim}15\%$ to ${\sim}45\%$, which is shown alongside literature values from previous multiplicity surveys in the bottom panel of Figure~\ref{fig:results}.

\section{Discussion} \label{sec:discussion}
The central observational result of this work is that the multiplicity fraction is approximately constant --- around $45\%$ --- across the primary-mass range $0.4-0.75\,M_\odot$. In contrast, there is a decline at the lowest M-dwarf masses to $\sim 15\%$ at $0.2\,M_\odot$.
This is not entirely aligned with the canonical picture of a monotonically rising mass--multiplicity relation \citep{DucheneKraus2013, MoeDiStefano2017, Offner+23}, which places the M-dwarf regime into a continuous decline towards the lowest stellar masses. Our results instead align the upper M-dwarf regime ($\gtrsim 0.4\,M_\odot$) with solar-type stars, with the multiplicity fraction only drop at lower masses.
That makes it important to discuss both the absolute MF values and the shape of this relation in the context of previous observations and binary-formation physics.

\subsection{Comparison with M-star multiplicity studies}
Previous companion-counting studies of M-dwarfs typically report multiplicity fractions near $25\%$ \citep[e.g.][]{WardDuong+15, Winters+19, Clark+24}, but \cite{Cifuentes+25} argue that this is an underestimate: once \tgaia-identified astrometric candidates are added in, the fraction rises to $40.3\%$ \citep{Cifuentes+25}, which our extrapolated MF across our sample is broadly consistent with.
Additionally, good agreement is found with hydrodynamical simulations for both the rates of multiplicity and the mass-dependent structure \citep{Bate2012}.
Finally, \cite{Gonzalez-Payo2026} compiled sources within 10\,pc and reported multiplicities as a function of primary mass which we broadly agree with, except at the lowest masses (Orange crosses, Figure~\ref{fig:results}), where we also disagree with the results from \cite{Winters+19}, and it warrants discussing why this may be the case.

We report global multiplicity fractions extrapolated across all mass ratios, excluding planetary companions, but we retain the contribution from brown dwarfs.
Different studies define multiplicity differently, making direct comparison challenging, however some do expressly include brown dwarfs in their rates of multiplicity
\citep{Gonzalez-Payo2026, Bate2012}, which we are in broad agreement with.
Since we only fit multiplicities for companions $q\lesssim0.35$ (Figure~\ref{fig:MassCutoff}), at $0.25\,M_\odot$ this places nearly all of our companions used at inference in the brown dwarf $M_2\lesssim80M_\text{Jup}$ regime, and therefore the fact that we disagree with \cite{Gonzalez-Payo2026} and \cite{Winters+19} here can either represent a true difference, or indicate that the fiducial models no longer hold into the brown dwarf regime.

Particularly, we assume a flat mass ratio distribution. But, if stellar companions are in fact far more likely than brown dwarf ones -- a ``Brown Dwarf Desert" \citep[e.g][]{GretnerLineweaver06, Kraus2008, Kiefer2019} -- then our under-prediction of the rates of multiplicity in these low mass bins is to be expected.
To align the lowest-mass bin with the previous studies, this would require stellar companions to be roughly $1.7$ times more likely than brown dwarf companions.

One caveat to the absolute MF values, and to their mass dependence specifically, deserves explicit mention.
Old, faint white-dwarf companions may pass our spectral analysis and isochrone cuts: they contribute negligible flux and do not shift the system off the main sequence. Yet, they produce strong astrometric signals and are counted as multiples in our inference, with their actual mass ratios ($q \gtrsim 1$) outside the $q \in (0.05, 0.5)$ grid and so implicitly absorbed into the stellar-companion component. The WD-companion fraction for M-dwarf primaries is at the percent level \citep{MoeDiStefano2017, Shahaf+2024}, and is expected to increase with primary mass because the higher luminosity and more blue intrinsic colour results in a larger fraction of WDs that can be hidden.
Additionally, \cite{MoeDiStefano2017} propose that unresolved WD companions have been not counted in the volume-limited multiplicity studies \citep[e.g.][]{Raghavan+2010}.
A small fraction of the rise we attribute to stellar-companion multiplicity may therefore in fact reflect a rising WD-companion contribution, especially in the highest mass bin, but would strengthen the apparent plateau shape above $0.4\,M_\odot$.

\subsection{The plateau} 
The total multiplicity plateau lower limit at $0.4-0.75\,M_\odot$ is around $45\%$, putting it in line with previous studies of multiplicity of solar type stars \citep{Raghavan+2010,Tokovinin14,MoeDiStefano2017}, which extends to ${\sim}1.5\,M_\odot$ \citep{Tokovinin14}.
If real, this plateau is the most surprising feature of our result, since it would imply that mass--multiplicity is approximately flat across more than a factor of three in primary mass.

Dynamical few-body interactions in young clusters tend to ionise weakly-bound higher order systems and eject the lowest-mass component, broadly decreasing multiplicity with time \citep{Heggie75, Hills75}.
Simulations of star forming regions generally find high initial multiplicity fractions before dynamical processing, and large stars have a very high multiplicity fraction partially because they don't live long enough to see the dynamical processing entirely relax them to a steady-state, and therefore multiplicity should decrease with primary mass \citep[e.g.][]{Kroupa95, Goodwin+07, Bate2012}.
An interpretation of this $0.4$--$1.5\,M_\odot$ multiplicity plateau is that stars of F-type or later are typically observed in the solar neighbourhood after dynamical processing is largely complete, since these stars are typically older than a few hundred million years, and dynamical processing settles at no longer than this timescale \citep[e.g. Figure 3 of][]{Goodwin+07}.

Stars forming with high rates of multiplicities which dynamically settle near $50\%$ in the solar- to M-type regimes not dissimilar to our results has been found in simulations of star forming clouds \citep{Bate2012}.
A steady state multiplicity fraction of $50\%$ can be expected at the end of dynamical processing if the typical final dynamical interaction that each system experiences results in an ejected component and a remaining binary (or a stable higher order system), as this corresponds to a multiplicity of $50\%$.
Surveys of young stars have generally revealed higher multiplicities \citep{Ghez1993, Kraus2011}, however it would remains prudent to make further observations of stellar populations of a variety of ages to empirically tie down this dependence.

\subsection{The drop at low-mass}
Star forming clouds are generally self-similar across mass scales \citep{Larson92}, but evidently there must be some lower mass limit at which the self similarity ceases due to the limiting star forming cloud size and efficiency \citep{Goodwin+07}.
The rate of companions declining in primaries below $0.4\,M_\odot$ could be indicative that the self-similarity of the star forming cloud changes, or forming companions becomes less efficient, at this mass.
It is also about at this mass where the IMF turns from a higher-mass power law down into something closer to a log-normal \citep[e.g.][]{Kroupa01, deMarchiParesce01, Chabrier05}, neatly summarised in Figure~1 of \citet{Offner+14}.

The location of our drop (${\sim}0.4\,M_\odot$) is intriguingly close to several stellar-physics transitions in the same mass range: this IMF turnover, the fully-convective boundary at ${\sim}0.35\,M_\odot$, and the opacity-limited fragmentation scale below which collapse fragments are increasingly hard to support \citep[e.g.][]{Bate2012}. 
Radiation-hydrodynamic simulations of cluster formation likewise predict structure in the multiplicity--mass relation in this regime \citep{Bate2012, KratterLodato2016}, although precise quantitative predictions for $0.2$--$0.75\,M_\odot$ primaries do not yet exist at the precision needed to compare with our observational result. 
We leave it to future work to investigate these courses of inquiry, and foresee that future data releases will provide the opportunity to scrutinise the interpretation of these results, as, for example, epoch astrometric data in \tgaia's next data release can reveal as of yet unknown systematics.

\section{Summary} \label{sec:summary}
We have demonstrated that the astrometric solution assignment in \tgaia's third data release is sensitive to the period, mass ratio, and eccentricity distribution of binaries, and that this dependency is predictable and can be accurately forward-modelled using the generative forward model {\tt Gaiamock} \citep{ElBadry+2024}.
We show how to give a robust and percent-level estimate of the multiplicity fraction for effectively invisible companions of a given sample, given fiducial choices for the distribution of the orbital parameters, and have validated the self-consistency of this method on synthetic data.
As a demonstration, we take a sample of $366~027$ primaries of $0.2$ to $0.75\,M_\odot$ between $50$ and $200$\,pc with spectroscopic analysis to eliminate companions contributing significantly to the flux, and apply strict cuts to eliminate spurious astrometric signals.
With these samples in hand, we determine the multiplicity fraction required to explain their sets of astrometric signals, and how this multiplicity fraction varies with primary mass.

\subsection{Mass-multiplicity relationship} Our central result is that the multiplicity fraction appears approximately constant around $45\%$ across the primary-mass range ${\sim}0.4$--$0.75\,M_\odot$ with a decrease to near ${\sim}15\%$ by $0.2\,M_\odot$ (Figure~\ref{fig:results}). 
The flat behaviour above $0.4\,M_\odot$ revises the canonical picture of a monotonically rising mass--multiplicity relation, aligning the upper M-dwarf regime with solar-type stars rather than placing it on a continuous decline towards the lowest stellar masses. 
The drop below ${\sim}0.4\,M_\odot$ coincides with the IMF turnover and the fully-convective boundary, suggesting a qualitative change in the dominant binary-formation channel at these masses. While the precise values of the multiplicity fractions are sensitive to modelling choices, the qualitative shape of the mass--multiplicity relation is robust to the tests we performed (Appendix~\ref{app:Robustness}) and is itself the most physically informative result.

\subsection{Global multiplicity fraction} Our extrapolated multiplicity fractions are higher than most previous M-dwarf studies, which place the multiplicity fraction near $25$--$30\%$ \citep[e.g.][]{WardDuong+15, Winters+19, Clark+24}, but we are in broad agreement with the higher values reported in \cite{Gonzalez-Payo2026}, and also with \cite{Cifuentes+25} who estimated a corrected fraction of $40.3\%$ across M-type primaries once unresolved binary candidates are accounted for. 

However, at the lowest mass end, we predict lower multiplicities than that in \cite{Gonzalez-Payo2026} and \cite{Winters+19}, which may be due to brown dwarf companions dominating our inference for these primaries, and this difference can be explained if brown dwarf companions are ${\sim}1.7$ times less likely than a stellar companion.

\subsection{Limitations in DR3} Our incomplete understanding of \tgaia's inner workings (here with respect to the increased errors and uncertainties in our sample of faint, red objects) limit our ability to derive orbital parameter constraints. 
In particular, our forward model over-predicts the rate of full orbit solutions in this regime relative to the observed sample (Section~\ref{sec:modelling}), pointing to systematics in the high-$\truwe$ tail that we do not fully reproduce. 
However, the \ruwe distribution can be well captured \citep{iorio+26}, which provides a degree of freedom which allows the multiplicity fraction to be precisely estimated, and we have demonstrated the ability to recover a mass-multiplicity relationship in synthetic data.
Additionally, while the inferred MF is model dependant, we have shown that the model dependence contributes but $\mathcal{O}(\%)$ biases (Appendix~\ref{app:Robustness}).
Epoch astrometric data, which will be available in \tgaia's next data release, will provide the requisite data to create forward models with the precision needed to use the rates of the higher order solution types to place constraints on the distributions of orbital parameters.
We also do not separately model compact-object (e.g.\ white-dwarf) companions, which in principle can pass our photometric cleaning and contribute at the percent level to the inferred MF; see Section~\ref{sec:results} for a fuller discussion.

This paper is a first demonstration of a forward-modelling framework that converts the otherwise-overlooked subset of \gaia astrometric diagnostics such as elevated \ruwe and accepted acceleration/jerk solutions into a population-level multiplicity constraint. 
We view its current scope as deliberately conservative: only the multiplicity fraction is fit, not the underlying distributions of $(T,e,q)$, because the absolute calibration of the higher-order solution rates is not yet trustworthy enough to support such constraints from DR3 alone.
However, we view sensitivity to orbital parameters not as a limitation, but rather as an opportunity: with future data releases, especially the per-epoch astrometric data anticipated in \tgaia's next release, this framework should be capable of jointly constraining rates of multiplicity and the distributions of binary parameters, with the absolute calibration tied directly to the raw data rather than to the published solution-type rates. 
We anxiously await these capabilities.

\begin{acknowledgements}
B.~Pennell, H.~W.~Rix, S.~Shahaf, J.~M\"uller-Horn, and J.~Li acknowledge support from the European Research Council for the ERC Advanced Grant [101054731]. 
This work has made use of data from the European Space Agency (ESA) mission \textit{Gaia} \url{https://www.cosmos.esa.int/gaia}, processed by the \textit{Gaia} Data Processing and Analysis Consortium (DPAC, \url{https://www.cosmos.esa.int/web/gaia/dpac/consortium}). Funding for the DPAC has been provided by national institutions, in particular the institutions participating in the \textit{Gaia} Multilateral Agreement. We would like to thank the anonymous referee for strengthening this paper. Absolutely no generative AI was used in the writing of this text.
\end{acknowledgements}

%
   \bibliographystyle{aa} 
   \bibliography{references.bib} 
%

\begin{appendix}

\nolinenumbers
\section{Sample creation} \label{app:Sample}
\begin{figure}
    \centering
    \includegraphics[width=\linewidth]{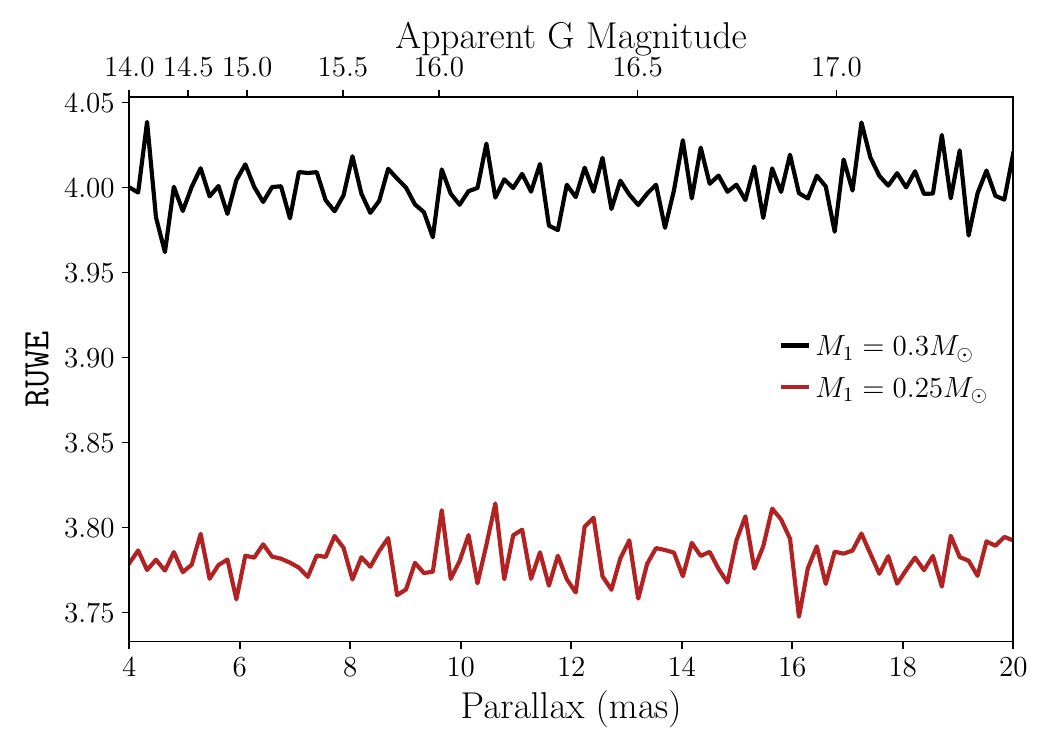}
    \caption{Demonstration that constant Astrometric Significance yields unchanging astrometric observations. We show \ruwe calculated by {\tt Gaiamock} for a $0.25\,M_\odot$ (red) and $0.3\,M_\odot$ (black) primary with a $0.1\,M_\odot$ companion on a $700$\,day period. The parallax was changed, and the magnitude was commensurably adjusted to maintain the same Astrometric Significance $\Lambda$ by Equation~\ref{eq:AstrometricSignificance}. The assigned \ruwe would typically increase with parallax, but we are able to assign the magnitude which would maintain the same \ruwe due to our derivation of $\Lambda$. Since the \ruwe is constant as a function of parallax and magnitude, we can compute \ruwe a single time and look it up for any other combination of parallax and magnitude (and primary and secondary mass) without needing to compute the astrometry again.}
    \label{fig:LambdaDemonstration}
\end{figure}

\subsection{Flux ratio limit} \label{app:fcut}
\begin{figure}
    \centering
    \includegraphics[width=\linewidth]{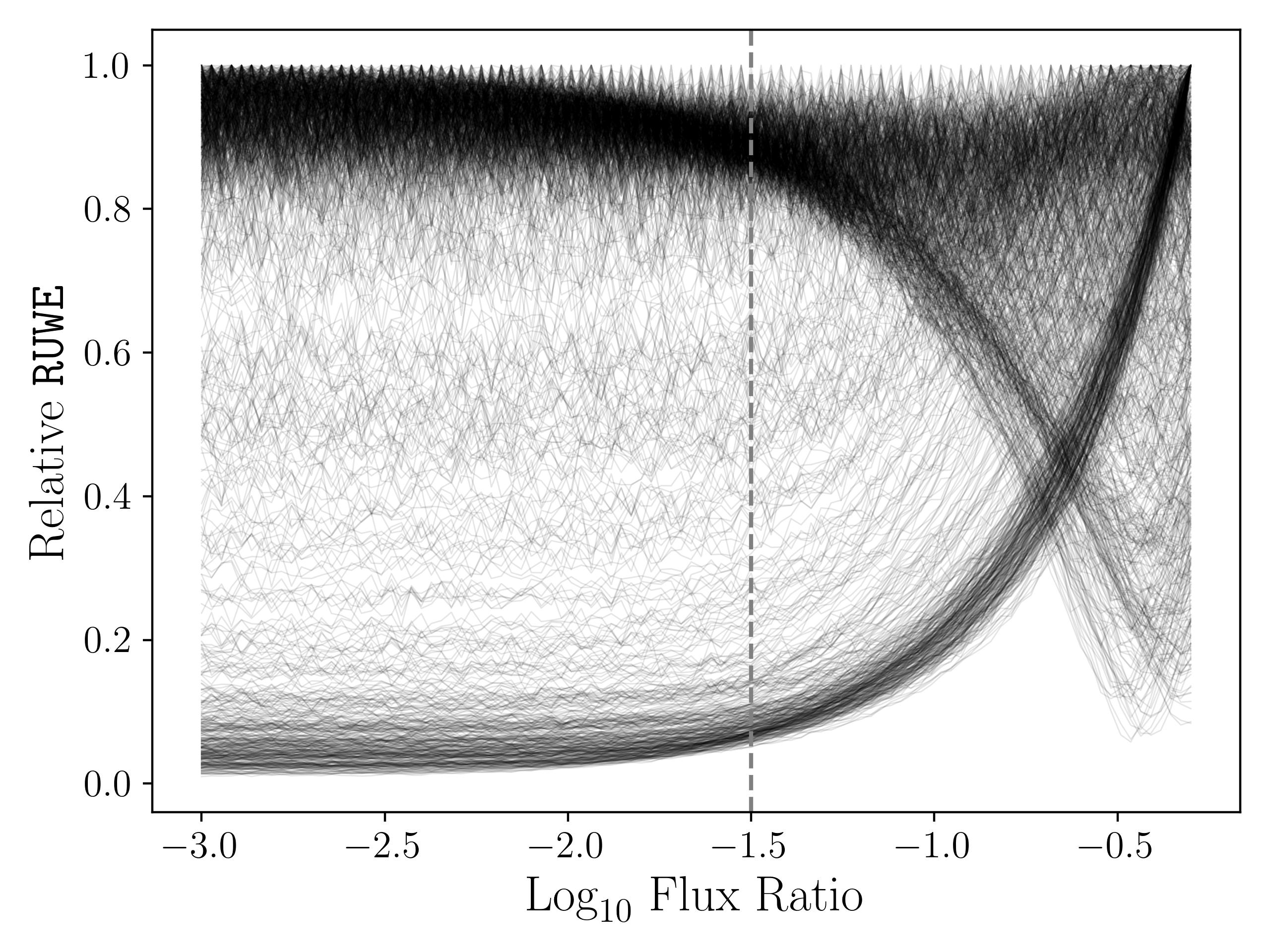}
    \caption{\ruwe simulated by {\tt Gaiamock} for binary systems as a function of flux ratio, normalised to the maximum \ruwe attained per object. This shows that below $f=10^{-1.5}$, the flux ratio contributes little to the astrometric signals.}
    \label{fig:fluxcut}
\end{figure}

In Section~\ref{sec:query}, we aimed to spectroscopically remove systems with companions significantly contributing to the flux through the spectral modelling from \cite{Li+2025}, by eliminating sources with binary models best fit with a high flux ratio.
We must choose a flux ratio to make this cut, and we aim to find a flux ratio below which {\tt Gaiamock} would not meaningfully assign different astrometry signals.
To determine this, we simulate many binaries at different flux ratios.
We simulate $1000$ systems with primary masses, sky positions, and proper motions sampled from our parent sample from Section~\ref{sec:query}, with randomly assigned companions over uniform priors $\bigp(\log{T})=\mathcal{U}(2,6)$, $\bigp(q)=\mathcal{U}(0.3,0.5)$, $\bigp(e)=\mathcal{U}(0,1)$. 
We then used {\tt Gaiamock} to compute \ruwe for these objects at several different flux ratios, which can be seen in Figure~\ref{fig:fluxcut}
We have found that below roughly $f=10^{-1.5}\!\approx\!3\%$, the computed value of \ruwe changes very little with flux ratio relative to a far fainter companion, suggesting the flux contribution is negligible on the astrometric measurements below this regime.
At inference, we assumed all companions contribute no flux.
If at inference we attempted to model the flux contribution of the companion, this would inflate the fit multiplicity fractions.

\begin{figure}
    \centering
    \includegraphics[width=\linewidth]{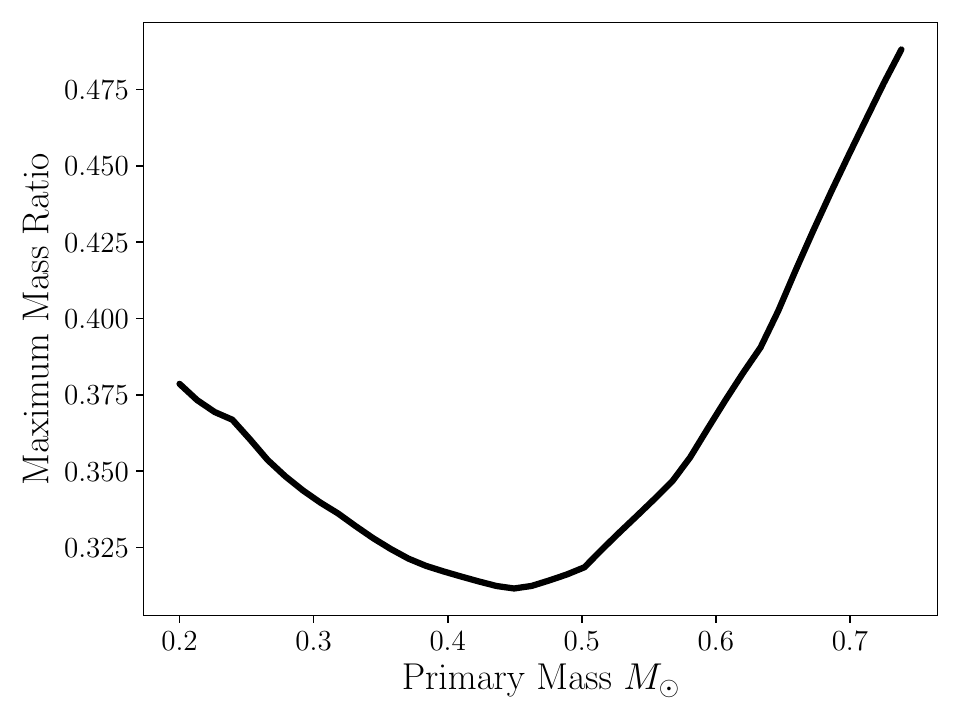}
    \caption{Mass ratios corresponding to a flux ratio of $10^{-1.5}$ for solar metallicity stars with an age of $10^9$ years as computed by PARSEC isochrones.}
    \label{fig:MassCutoff}
\end{figure}

We convert the adopted flux-ratio threshold into a mass-ratio limit using 1\,Gyr PARSEC isochrones \citep{Bressan2012}: for each primary mass and metallicity, we interpolate the primary $G$-band magnitude along the isochrone, imposing the change in magnitude due to the companion at a flux ratio $f=10^{-1.5}$, $\Delta G=-2.5\log_{10}(10^{-1.5})$\,mag, and interpolate to find secondary mass that would contribute this flux. 
This gives $q_{\rm cut}=M_2/M_1\sim0.3$--$0.4$ over the relevant mass and metallicity range, so our sample is mainly sensitive to companions with $q\lesssim0.35$ (Figure~\ref{fig:MassCutoff}).

\subsection{Identifying and removing spurious high \ruwe sources} \label{app:Spurious}
\begin{figure}
    \centering
    \includegraphics[width=\linewidth]{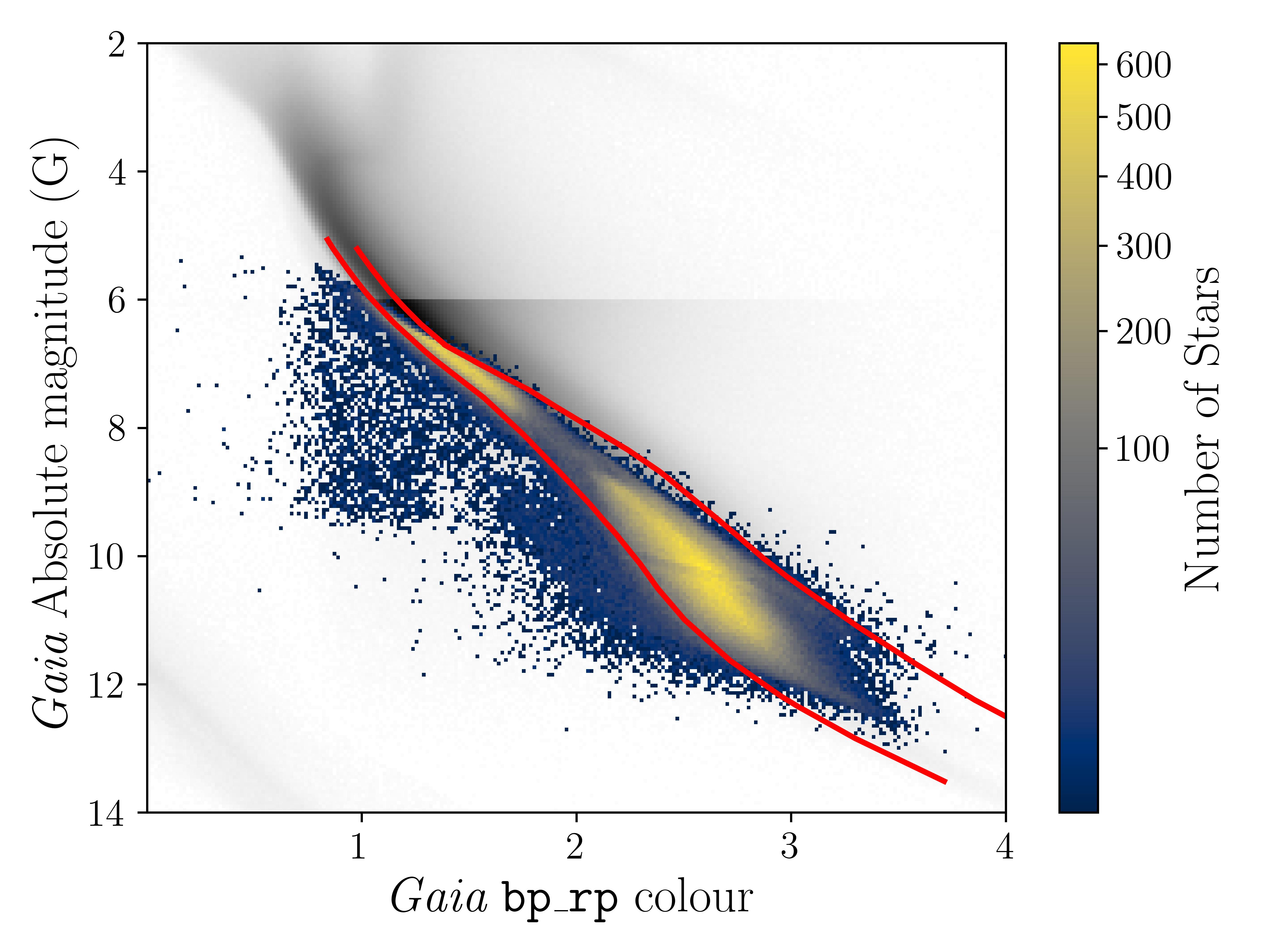}
    \caption{CMD density for the sample outlined in Section~\ref{sec:query}, with the parent sample showing the whole main sequence in grey, and the bounding isochrones in red, with objects outside these isochrones being thrown out, corresponding to $6\%$ of the sample.}
    \label{fig:isochronecuts}
\end{figure}

\begin{table*}[t!]
    \centering
    \caption{Source count and distributions for each cut in Table~\ref{tab:SelectionCuts}.}
    \begin{tabular}{ll|lllll}
    \hline
    \hline
    & & $\truwe\!<\!1.4$ & $\truwe\!>\!1.4$ & Accel. & Jerk & Full \\
    \hline
    Initial Cuts (Section~\ref{sec:query}) & Rate (\%) & 92.55 & 6.21 & 0.47 & 0.32 & 0.45 \\
                 & Count & 419597 & 28136 & 2142 & 1448 & 2042 \\
    \hline
         Isochrone Cut (Figure~\ref{fig:isochronecuts}) & Count & 20439 & 6129 & 106 & 127 & 254 \\ 
         Photometric variability & Count & 29217 & 2328 & 256 & 193 & 256 \\ 
         $\texttt{ipd\_frac\_multi\_peak} > 1$ & Count & 16357 & 16317 & 34 & 18 & 27 \\ 
         All three above cuts & Count & 63485 & 18958 & 379 & 319 & 496 \\ 
    \hline
        High $\delta g$ (Figure~\ref{fig:FluxError}) & Count & 1791 & 1881 & 14 & 8 & 6 \\ 
    \hline
         Final Catalogue & Rate (\%) & 96.80 & 1.99 & 0.48 & 0.31 & 0.42 \\
                & Count & 354320 & 7296 & 1749 & 1121 & 1540 \\
    \hline
    \end{tabular}
    \label{tab:SpuriousCuts}
\end{table*}

We don't have a complete understanding of the contamination of spurious objects getting a \ruwe above $1.4$. 
Under normal conditions, a veritable single star is not able to create a \ruwe this high \citep{CastroGinard+2024}, but the astrometric fits can be spurious due to variability, blending with nearby on-sky sources, or partially resolved binaries \citep[e.g.][]{GCNS}. 
Every object with spurious astrometry of this variety gets an inflated \ruwe but will rarely get a higher order solution assigned. 
For the closest 100\,pc objects, single stars get spurious solutions at a rate of ${\sim}0.15\%$, and it is foreseen that this rate should increase with distance \citep{GCNS}. 
While a $\mathcal{O}(0.1\%)$ spuriousness rate appears low, it could inflate fit multiplicity fractions by several percent. 
Additionally, any objects with bright companions which are identified as likely single by \cite{Li+2025} would inflate our fit multiplicity fraction.
We make four cuts to eliminate these sources which are either truly spurious or otherwise cannot be reproduced in the forward model, the summary of sources removed in each cut can be found in Table~\ref{tab:SpuriousCuts}.
Crucially, we aim to provide lower limits on the multiplicities, and the multiplicity is tied to the rates of high \truwe. 
Each following cuts are not only mindful but indeed remove a significant number of high \ruwe sources, thereby decreasing the inferred multiplicities.

We first compute isochrones bounding the main sequence above and below, with a hope to catch objects either with bright stellar companions, WD companions, or poor astrometric fits which assigned poor parallaxes, each of which can divert a regular star away from its expected position on the CMD.
We bound the main sequence with PARSEC isochrones \citep{Bressan2012} from below with a low metallicity of $\text{[M/H]}=-0.3$ and high age of $10^{10}$\,yr, and from above by a high metallicity $\text{[M/H]}=0.3$ and low age $10^{7.8}$\,yr, keeping only objects which fall between these isochrones. These isochrones overlaid on the CMD can be seen in Figure~\ref{fig:isochronecuts}.

The vast majority of effectively single stars in the solar neighbourhood should not fall outside these isochrones under normal conditions, so it is a suitable way to remove objects that are either spurious or have white dwarf or other bright companions that change the colour or magnitude off the main sequence. 

\begin{figure*}[t]
    \centering
    \includegraphics[width=\linewidth]{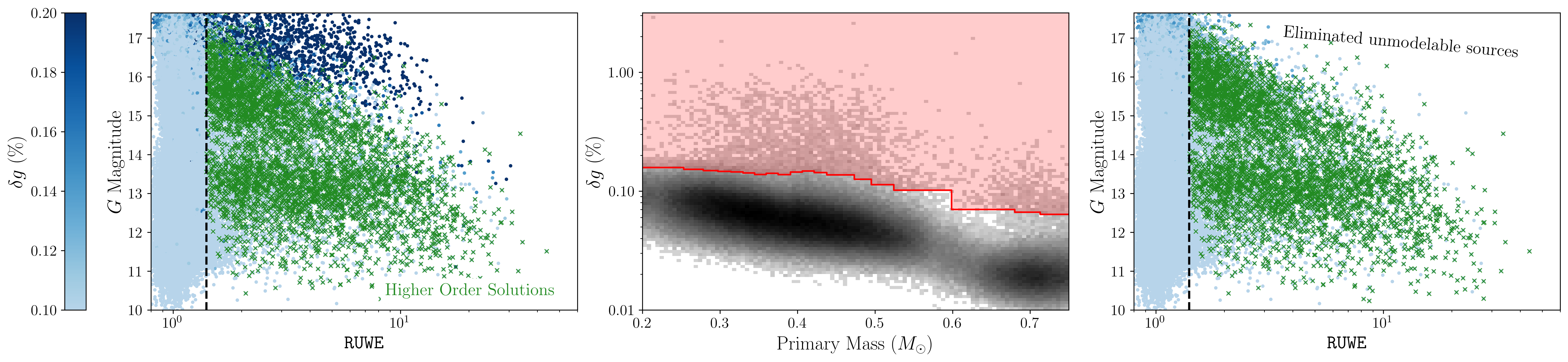}
    \caption{Demonstration of motivation for the fractional flux error ($\delta g = \texttt{phot\_g\_mean\_flux\_error}/\texttt{phot\_g\_mean\_flux}$) cut to address spurious \ruwe detections due to on-sky blending, with the sample shown before (left) and after (right) the $\delta g$ cut (middle). Blending inflates the flux error as background sources are inconsistently measured alongside the target by \tgaia, while this inconsistent blending makes them unlikely to be well fit by the higher order solutions. Left panel: distributions of \ruwe and apparent magnitude for the sample (Section~\ref{sec:query}) coloured by $\delta g$, with the higher order solutions shown in green. This shows a space where sources get high \ruwe values but do not get assigned higher order solutions (dark blue), which also get high flux errors $\delta g$, which indicates blending. Middle panel: density of sources in primary mass and $\delta g$. The sample is divided into mass bins, and the top $1\%$ of sources in $\delta g$ are removed from each bin, with this cut shown in red. Right panel: same as the left panel, but with the objects in the red region of the middle panel eliminated. This shows the efficacy of the cut in removing the objects in the region of interest, while minimally affecting the rest of the sources.}
    \label{fig:FluxError}
\end{figure*}

\begin{figure}
    \centering
    \includegraphics[width=\linewidth]{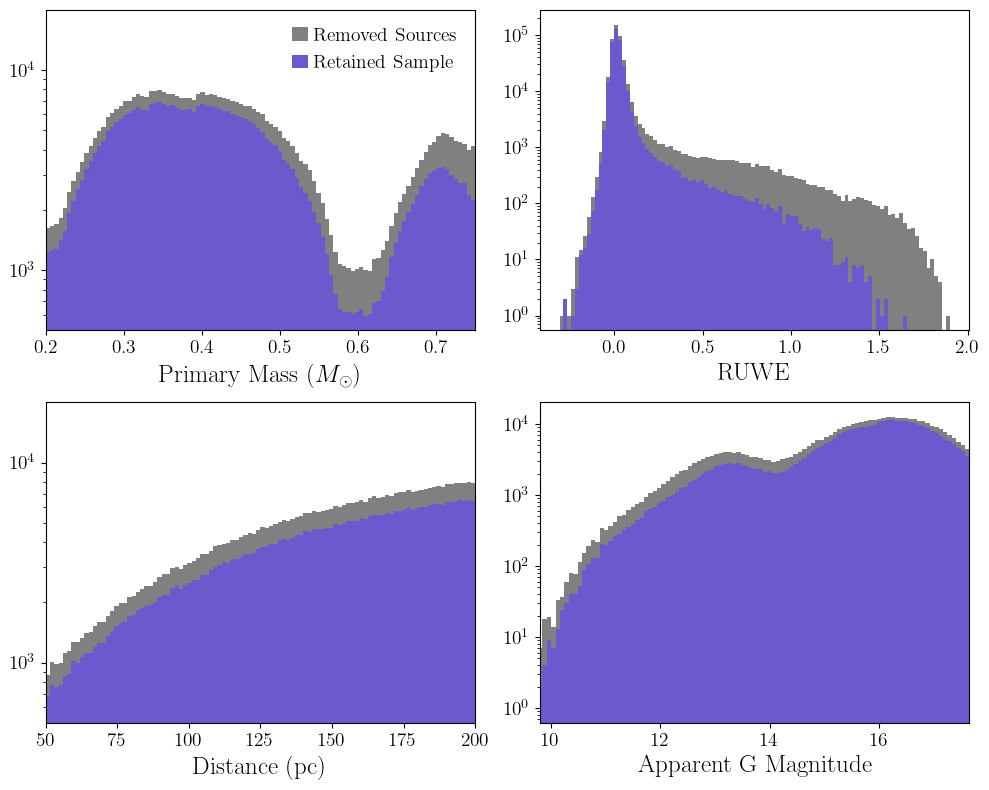}
    \caption{Distribution of primary mass, {\tt RUWE}, distance, and apparent magnitude in the final sample and the distribution of sources eliminated by the cuts in Table~\ref{tab:SpuriousCuts}.}
    \label{fig:CutObjects}
\end{figure}

Second, each source is published with a photometric variability flag $\texttt{phot\_variability\_flag}$.
Sources with high variability can affect the measured astrometry and should be removed from the sample.
Third, we eliminate sources with $\texttt{ipd\_frac\_multi\_peak}>1$, which indicates that there are two or more sources being blended together in the \gaia transits.
all three of these cuts have an inflated rate of high \ruwe (Table~\ref{tab:SpuriousCuts}) and therefore eliminating them reduced the fit multiplicities such that we report lower bounds.

Finally, we attempt to address objects whose \ruwe is inflated due to blending of nearby sources \citep{GCNS}. 
We have the advantage that objects which get inflated \ruwe due to blending would be unlikely to be well fit by a higher order model, since blending would be applied fairly randomly, so we would expect to see sources with high \ruwe but no acceleration, jerk, or orbit solutions assigned.
In the left panel of Figure~\ref{fig:FluxError}, we show the distribution of \ruwe and apparent magnitude of the remaining sample which shows that the dimmest and highest \ruwe sources do not get assigned higher order solution types.
Additionally, these sources cannot be generated by {\tt Gaiamock}.
This means that these objects are likely spurious, perhaps by blending, and either way are not accounted for in our forward model and should therefore be discarded.

Sources which have inflated \ruwe due to on-sky blending would likely have an inflated uncertainty in their magnitude, since definitionally they were scanned inconsistently with and without the interloping background source or companion.
Fortunately, the mean flux and flux error is published by \gaia for all sources, making this an easy-to-apply metric.
The sources which exist in the left panel of Figure~\ref{fig:FluxError} with high \ruwe but no higher order solution types also generally have an inflated fractional flux error $\delta g = \texttt{phot\_g\_mean\_flux\_error}/\texttt{phot\_g\_mean\_flux}$.

We aim to remove objects with fractional flux errors that are likely to occur only due to blending or some other observational error, by quantifying the reasonable upper limit on $\delta g$ for true single stars and binaries. 
We divide the sample into 20 equal sized bins by primary mass, and eliminate the sources with the highest $1\%$ in $\delta g$, which was found to optimise eliminating a high number of high \ruwe sources while retaining sources with higher order solutions, which suggests this cut is working as intended (Table~\ref{tab:SpuriousCuts}).

Overall, these three cuts eliminated the majority of the high \ruwe sources (Table~\ref{tab:SpuriousCuts}), which commensurably decreases the inferred multiplicity fractions. 
This shows the importance of making such cuts in this kind of inference.
This further highlights the role that epoch astrometry will play in \tgaia's next data release, as we will have a better ability to incorporate features such as blending through outlier treatments into our modelling.
We acknowledge the sensitivity of our results to such sample creation, however these cuts rather homogeneously eliminate sources (Figure~\ref{fig:CutObjects}).
For this reason these cuts do not significantly alter the shape of the mass-multiplicity relationship in Figure~\ref{fig:results}.
Finally, since we've made conservative choices so as to lower the multiplicity fractions, the multiplicities we present can be interpreted as lower limits. 

\subsection{Common proper motion pairs} \label{app:cpm}
\begin{figure}
    \centering
    \includegraphics[width=\linewidth]{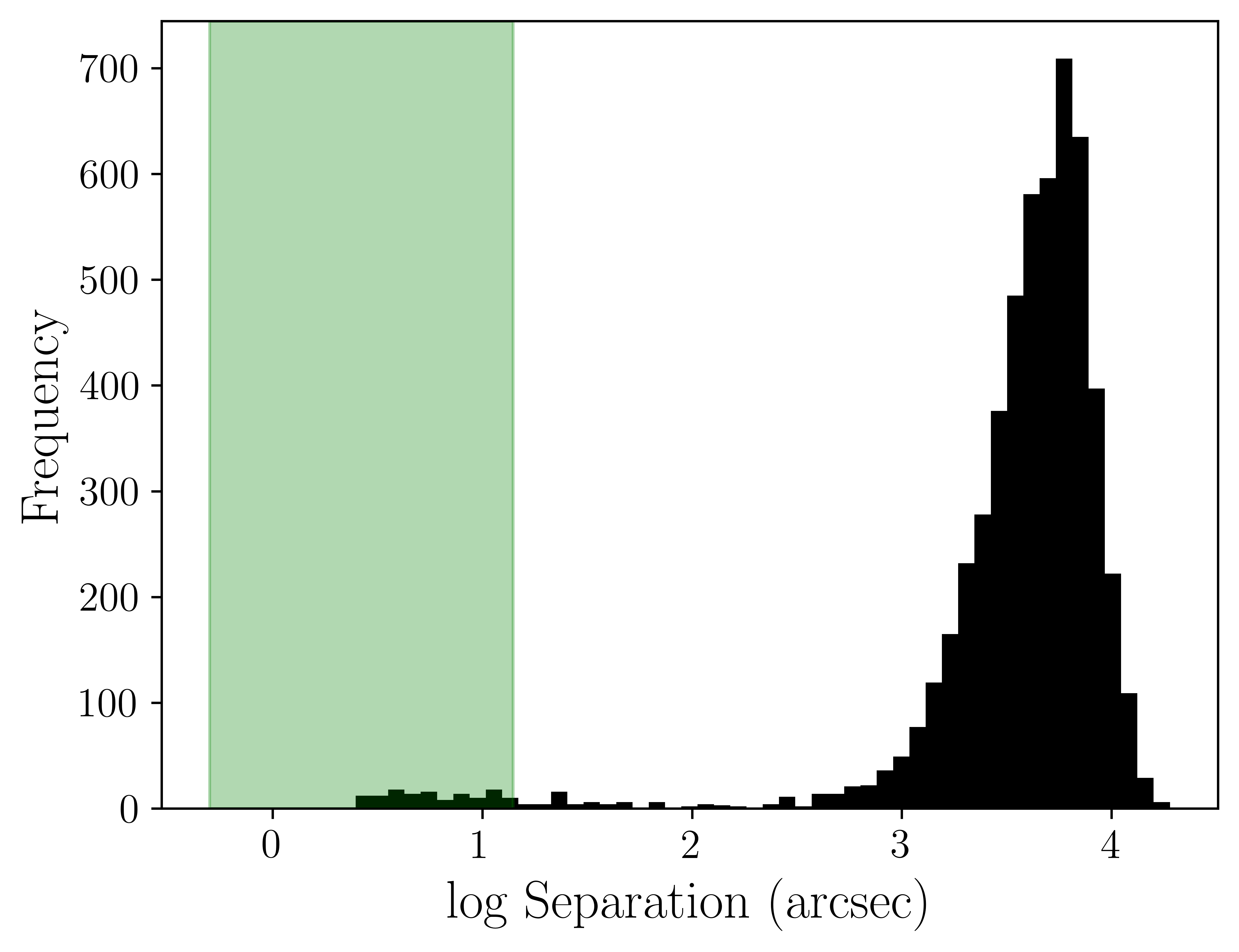}
    \caption{Histogram of the angular separation to nearest neighbour for sources between 50 and 60\,pc in the final sample, with the reasonable angular separation of wide binaries in green. Demonstrates how there is a very small contamination of resolved binaries in the sample.}
    \label{fig:widebinaries}
\end{figure}

A potential contaminant in the sample is wide binaries which are sufficiently separated that \gaia would assign each component a separate source id.
This would cause two apparently single stars to enter the sample, when they should be a single binary star.
Many systems near the resolving limit will have been removed by the cuts in the previous section, however, in principle, entirely resolved pairs can remain.

We only took sources beyond 50\,pc to account for this, since this would reduce the possible parameter space of these wide companions.
In Figure~\ref{fig:widebinaries}, we took all sources in our final sample between 50 and 60\,pc and determined their nearest angular neighbour. 
A wide binary would be expected to be far beyond \tgaia's diffraction limit of $\mathcal{O}(0.1")$ \citep{DR3}, but close enough to be bound.
In Figure~\ref{fig:widebinaries} we shade in green the region from $0.5"$ to the separation of a companion to an M-type primary with an orbital period of $10^7$\,days, which is around $14"$.
As can be seen, there are remarkably few systems which fall in this range, which confirms that these wide binaries will contribute negligibly to the fit multiplicities.

\section{Luminous companions} \label{app:LuminousCompanions}
In Section~\ref{sec:Wrangling}, we took the assumption that the companion is invisible, and computed the astrometric amplitude parameter, $\lambda$, through which the astrometry scaled. We can generalise to a light contribution companion just as was done with the Astrometric Mass-Ratio Function in \cite{Shahaf+19}. 
For a luminous companion, the light centroid orbit size is the difference between the size of the orbit of the primary\footnote{You'll get the same answer if you do this computation using the secondary object.} around the centre of mass $a_{m,1}$ and centre of light $a_{l,1}$,
\begin{equation}
    a_\text{phot} = a_{m,1} - a_{l,1}.
\end{equation}

The total orbit size is related to the orbit size around the centres of mass and light,
\begin{equation}
    a_\text{tot} = \frac{1 + q}{q} a_{m,1} = \frac{1 + f}{f} a_{l,1}.
\end{equation}
Which yields,
\begin{equation}
    a_\text{phot} = \biggl(\frac{q}{1+q} - \frac{f}{1+f}\biggr)~a_\text{tot}.
\end{equation}
A modified version of Equation~\ref{eq:a_tot} can be written, and the remaining logic to derive $\lambda$ unchanged.
In the simplest case where the mass-luminosity relationship is a power law $L\!\propto\!M^\gamma$, then the flux ratio would be a function only of the mass ratio $f\!=\!q^\gamma$, which would maintain the original dimensionality. This yields the orbit's astrometric amplitude parameter
\begin{equation}
    \lambda = \plx M_1^{1/3}\frac{(q-f)}{(1+f)(1+q)^{2/3}} 
            = \plx M_1^{1/3}\frac{(q-q^\gamma)}{(1+q^\gamma)(1+q)^{2/3}}.
\end{equation}
Note that this equation has the correct asymptotic behaviour: as $q\!\rightarrow\!f$, the centre of mass and centre of light coincide, generating zero astrometric motion. 

\section{Model choice sensitivity} \label{app:Robustness}
\begin{figure*}
    \centering
    \includegraphics[width=\textwidth]{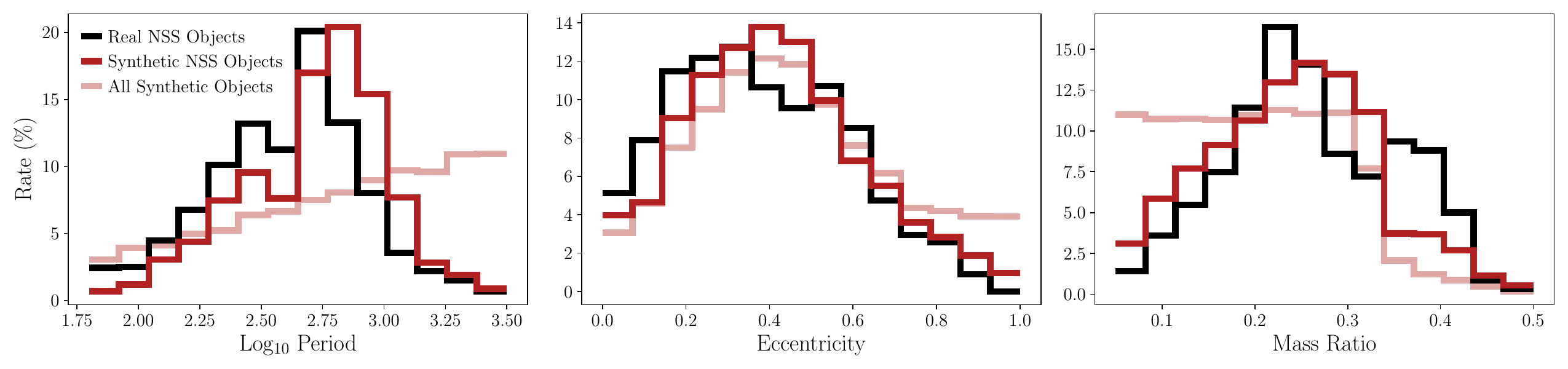}
    \caption{Comparison between real orbit solutions of nearby $0.2-0.75M_\odot$ primaries (black), the simulated objects (faint red) with a flat mass ratio distribution and log-normal period distribution peaking at $10^4$ days, and the simulated objects assigned orbit solutions (red). The rates across the relevant orbital parameters: period, eccentricity, and mass ratio, are shown. This shows that the fiducial models well reproduce the distribution of orbital parameters in the orbit solutions.}
    \label{fig:NSSComparison}
\end{figure*}
To demonstrate the robust nature of this constraining approach, we confirm that the multiplicity fraction can be recovered across several different synthetic datasets that span physically plausible models.

\begin{figure}
    \centering
    \includegraphics[width=\linewidth]{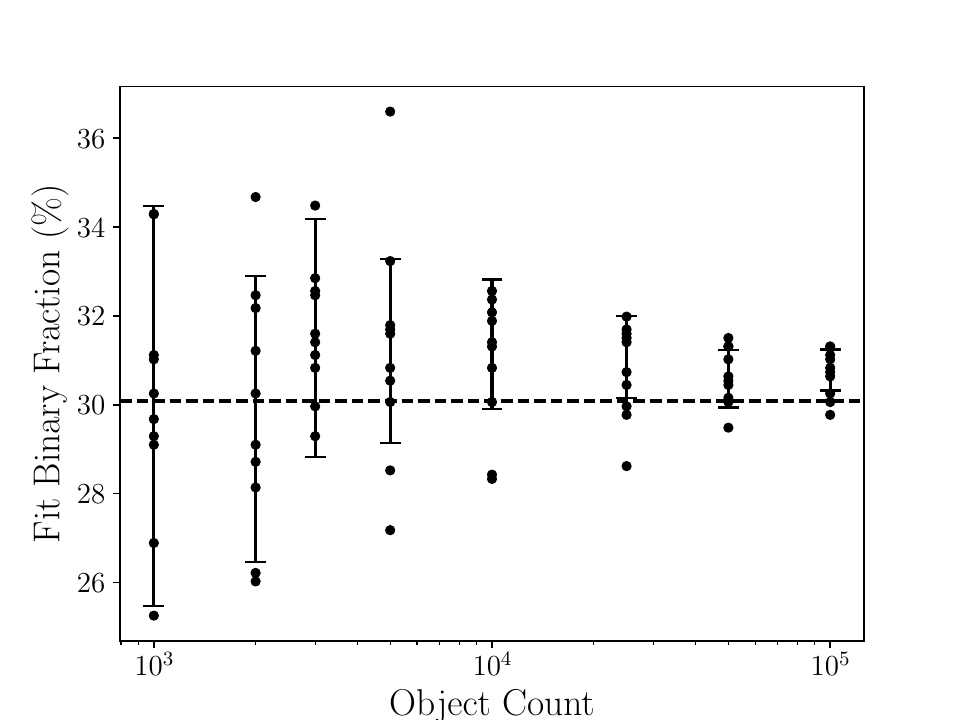}
    \caption{Test of the dependence of the number of objects on the constraining power for a sample with a global binary fraction of $30\%$. For each desired sample size, objects were sampled ten times from this synthetic dataset with replacement. The resulting fit multiplicity fraction for each sample is plotted in black circles, with the average Cramér-Rao lower bound uncertainty  represented as error bars emanating from the median fit multiplicity fraction. A small, sub-percent bias can be seen as the number of objects gets large.}
    \label{fig:CountDependence}
\end{figure}

We first determine the number of objects needed to place meaningful constraints. We generate samples with the same input distributions as Table~\ref{tab:choices} with an increasing number of objects. Figure~\ref{fig:CountDependence} shows how the uncertainties rapidly decrease up to $\mathcal{O}(10^4)$ objects to $\mathcal{O}(\%)$, and only modestly decrease beyond this. For this reason, a sample of more than $10^4$ objects allows percent-level constraints to be placed. Notably, as the number of objects well exceeds $10^5$, there is a consistent, sub-percent bias (a shift of about ${\sim}+0.7\%$, for this example), suggesting at this point unaccounted for systematics dominate the error, and therefore using more objects does not necessarily provide a better fit. 
Additionally, we place error bars corresponding to the Cramér-Rao lower bound, an approximation for the information contained. The spread of the bootstrapped fits shows that this is a good representation of the real inference uncertainty.

Next, we demonstrate the influence of the choice of period, mass ratio, and eccentricity distributions on synthetic samples created with $50~000$ objects.
For period distributions, we test several different log-normal distributions and the ability to recover a multiplicity fraction for synthetic data crafted with them. In addition to the suggestion from \cite{DucheneKraus2013} with $T_\mu=4$ and $T_\sigma=1.3$, we also test a model with $T_\mu=4$ and $T_\sigma=2$, and another with $T_\mu=5$ and $T_\sigma=1.3$. 
In Table~\ref{tab:PeriodDependence}, we show that a parent multiplicity fraction is recovered when the correct period distribution is used in the top three rows.
The recovered multiplicity fraction predictably rises or falls significantly when the wrong period distribution is employed, leading to errors of $\mathcal{O}(10\%)$, cautioning mindful choices for the period distribution, and the interpretation of the fit multiplicity fraction across all periods.
However, in the bottom three rows we convert the recovered multiplicity fractions into the multiplicity fraction of companions between $10^2$ and $10^4$ days and find they don't vary more than $\mathcal{O}(\%)$, which shows that in this regime where we're more sensitive to astrometry, the choice of period distribution provides comparatively small biases.

\begin{table}[ht!]
    \caption{Sensitivity of inferred multiplicity fractions due to orbital period distribution.}
    \centering
    \begin{tabular}{l|rrr}
        \hline
        \hline
         $\bigp(T)$ & $\text{MF} (4,1.3)$ & $\text{MF}(4,2)$ & $\text{MF}(5,1.3) $\\
        \hline
        (4, 1.3) & $\diagcell{30.64\!\pm\!2.11\%}$ & $32.85\!\pm\!2.50\%$ & $64.94\!\pm\!4.80\%$ \\
        (4, 2) & $26.50\!\pm\!2.02\%$ & $\diagcell{29.58\!\pm\!2.31\%}$ & $60.91\!\pm\!4.61\%$ \\
        (5, 1.3) & $13.34\!\pm\!1.44\%$ & $17.09\!\pm\!1.73\%$ & $\diagcell{29.29\!\pm\!3.27\%}$ \\
        \hline
        (4, 1.3) & $13.58\!\pm\!0.47\%$ & $12.32\!\pm\!0.44\%$ & $13.82\!\pm\!0.49\%$ \\
        (4, 2) & $11.75\!\pm\!0.44\%$ & $11.09\!\pm\!0.42\%$ & $12.96\!\pm\!0.48\%$ \\
        (5, 1.3) & $5.91\!\pm\!0.31\%$ & $6.41\!\pm\!0.32\%$ & $6.23\!\pm\!0.33\%$ \\
        \hline
    \end{tabular}
    \label{tab:PeriodDependence}
\end{table}

Likewise, we can recover the multiplicity fraction for synthetic datasets constructed with different mass ratio distributions (Table~\ref{tab:MassRatioDependence}).
We generate three synthetic catalogues with $50~000$ sources with a power law mass ratio distribution $\bigp(q)=q^\gamma$ with indices $\gamma=0,0.5,1$.
As with the period distributions, we also compare the recovered multiplicity when the wrong model is employed, which for the choice of mass ratio only changes the result by a $\mathcal{O}(\%)$.

\begin{table}[ht!]
    \caption{Sensitivity of inferred multiplicity fractions due to mass ratio distribution.}
    \centering
    \begin{tabular}{l|rrr}
        \hline
        \hline
         $\bigp(q)=q^\gamma$ & $\text{MF}(\gamma=0)$ & $\text{MF}(\gamma=0.5)$ & $\text{MF}(\gamma=1)$ \\
        \hline        
        $\gamma=0$ & $\diagcell{30.73\!\pm\!1.13\%}$ & $27.85\!\pm\!1.01\%$ & $24.77\!\pm\!0.91\%$ \\
        $\gamma=0.5$ & $34.00\!\pm\!1.19\%$ & $\diagcell{30.35\!\pm\!1.05\%}$ & $26.50\!\pm\!0.94\%$ \\
        $\gamma=1$ & $37.75\!\pm\!1.24\%$ & $32.46\!\pm\!1.08\%$ & $\diagcell{30.16\!\pm\!1.00\%}$ \\
        \hline
    \end{tabular}
    \label{tab:MassRatioDependence}
\end{table}

Finally, we employ the three eccentricity distributions outlined in Section~\ref{sec:constraining} to discern the sensitivity. Synthetic datasets of $50~000$ sources were created with circular $e\equiv0$, the thermal $\bigp(e)\propto e$ and the Rayleigh-thermal `turnover' model (Equation~\ref{eq:eccentricity}). Table~\ref{tab:EccentricityDependence} shows how multiplicity fractions can be recovered for each model, and using the incorrect model leads to modest $\mathcal{O}(\%)$ biases.
This is particularly notable for the case of the thermal distribution and our `turnover' distribution, since if the binaries are not veritably thermal, one would bias their estimates of the multiplicity fraction by blindly using that model.
For this reason, we caution the reader to carefully consider their choice of eccentricity distribution.

\begin{table}[ht!]
    \caption{Sensitivity of inferred multiplicity fractions due to eccentricity distribution.}
    \centering
    \begin{tabular}{l|rrr}
        \hline
        \hline
        $\bigp(e)$ & $\text{MF}(\text{Circular})$ & $\text{MF}(\text{Turnover})$ & $\text{MF}(\text{Thermal})$ \\
        \hline
        Circular & $\diagcell{30.54\!\pm\!0.84\%}$ & $31.41\!\pm\!0.87\%$ & $36.98\!\pm\!1.00\%$ \\
        Turnover & $28.71\!\pm\!0.82\%$ & $\diagcell{29.87\!\pm\!0.85\%}$ & $33.04\!\pm\!0.95\%$ \\
        Thermal & $25.74\!\pm\!0.78\%$ & $28.71\!\pm\!0.84\%$ & $\diagcell{32.08\!\pm\!0.94\%}$ \\
        \hline
    \end{tabular}
    \label{tab:EccentricityDependence}
\end{table}

\subsection{Fiducial model orbit solution comparison} \label{app:NSSComparison}
We compare our fiducial models to the observed sample of objects with orbit solutions. 
Our sample contained 1540 sources with full orbit solutions published in the \gaia DR3 NSS catalogue.
Assuming the companions are effectively invisible, we can compute the companion mass from the orbit parameters provided.

Next, using the fiducial parameter distributions described in Table~\ref{tab:choices} and scheme described in Section~\ref{sec:synthdata}, we simulated $50~000$ binaries and used {\tt Gaiamock} to propagate them through the \gaia astrometric pipeline.
This resulted in many synthetic orbit solutions, which we compare to the observed sample in Figure~\ref{fig:NSSComparison}.
This shows that the fiducial models are reasonable matches to the observed sample in the period range where orbit solutions are found.
Notably, the eccentricity distribution shows a very good match.
The mass ratio distributions are slightly misaligned, however, imposing a mass ratio distribution weighted to lower mass companions would increase the fit multiplicity fraction, which enforces our reported values as lower bounds.

\end{appendix}

\end{document}